\documentclass{article}
\usepackage{times}
\usepackage{epsfig}
\usepackage{graphicx}
\usepackage{amsmath}
\usepackage{amssymb}
\usepackage{helvet} % DO NOT CHANGE THIS
\usepackage{courier}  % DO NOT CHANGE THIS
\usepackage{graphicx}  % DO NOT CHANGE THIS
\usepackage{epsfig}
\usepackage{amsmath}
\usepackage{amssymb}
\usepackage{bm}
\usepackage{subfigure}
\usepackage{booktabs}
\usepackage{multirow}
\usepackage{booktabs}
\usepackage{caption}
\usepackage{float}
\usepackage{algorithmic}
\usepackage{algorithm}
\usepackage{bbding}
\usepackage{array}
\usepackage{color}
\usepackage{diagbox}
\usepackage{comment}

\floatname{algorithm}{Procedure}

\usepackage[square,sort,comma,numbers]{natbib}

\usepackage[final]{neurips_2020}

\usepackage[utf8]{inputenc} % allow utf-8 input
\usepackage[T1]{fontenc}    % use 8-bit T1 fonts
\usepackage{hyperref}       % hyperlinks
\usepackage{url}            % simple URL typesetting
\usepackage{booktabs}       % professional-quality tables
\usepackage{amsfonts}       % blackboard math symbols
\usepackage{nicefrac}       % compact symbols for 1/2, etc.
\usepackage{microtype}      % microtypography

\title{Self-Adaptively Learning to Demoir\'{e} from Focused and Defocused Image Pairs }

\author{
Lin Liu\textsuperscript{1,2} \hspace{1mm}
Shanxin Yuan\textsuperscript{2}\thanks{Corresponding author} \hspace{1mm}
Jianzhuang Liu\textsuperscript{2}\hspace{1mm}
Liping Bao\textsuperscript{1} \hspace{1mm}
Gregory Slabaugh\textsuperscript{2} \hspace{1mm}
Qi Tian\textsuperscript{3} \\ \\
 \footnotesize{$^1$EEIS Department, University of Science and Technology of China}\\ \footnotesize{$^2$Noah's Ark Lab, Huawei Technologies \qquad $^3$Huawei Cloud BU} \\ 
 \scriptsize{
   \{ll0825,baoliping\}mail.ustc.edu.cn  \{shanxin.yuan, liu.jianzhuang, gregory.slabaugh, tian.qi1\}@huawei.com}
}

\begin{document}

\maketitle

\begin{abstract}
%problem
Moir\'{e} artifacts are common in digital photography, resulting from the interference between high-frequency scene content and the color filter array of the camera.
Existing deep learning-based demoir\'{e}ing methods trained on large scale datasets are limited in handling various complex moir\'{e} patterns, and mainly focus on demoir\'{e}ing of photos taken of digital displays.
Moreover, obtaining moir\'{e}-free ground-truth in natural scenes is difficult but needed for training. 
In this paper, we propose a self-adaptive learning method for demoir\'{e}ing a high-frequency image, with the help of an additional defocused moir\'{e}-free blur image. 
Given an image degraded with moir\'{e} artifacts and a moir\'{e}-free blur image, our network predicts a moir\'{e}-free clean image and a blur kernel with a self-adaptive strategy that does not require an explicit training stage, instead performing test-time adaptation.
Our model has two sub-networks and works iteratively. During each iteration, one sub-network takes the moir\'{e} image as input, removing moiré patterns and restoring image details, and the other sub-network estimates the blur kernel from the blur image. The two sub-networks are jointly optimized.
Extensive experiments demonstrate that our method outperforms state-of-the-art methods and can produce high-quality demoir\'{e}d results. 
It can generalize well to the task of removing moir\'{e} artifacts caused by display screens. 
In addition, we build a new moir\'{e} dataset, including images with screen and texture moir\'{e} artifacts. As far as we know, this is the first dataset with real texture moir\'{e} patterns. 

\end{abstract}

\section{Introduction}

Image demoir\'{e}ing is the task of removing moir\'{e} patterns from images, taken by digital cameras from screens or from natural images with high-frequency patterns. 
Moir\'{e} artifacts are caused by the interference between the color filter array (CFA) of a camera and high-frequency repetitive signals, which can result from an LCD screen's subpixel layout or a natural scene's high-frequency repetitive patterns (e.g., textures on clothes).
Image demoir\'{e}ing is challenging as the moir\'{e} patterns vary in shape, color, and frequency. 
Existing deep learning based demoir\'{e}ing models \cite{sun2018moire,liu2018demoir, he2019mop,zheng2020image,liu2020wavelet} rely heavily on training with large amounts of annotated clean and moir\'{e} image pairs in order to obtain good performance. However, the models are still limited in handling various complex moir\'{e} patterns. Moreover, these models are restricted to perform demoir\'{e}ing of images captured from screens, having difficulty in removing moir\'{e} artifacts from natural images.

Lack of high-quality training data also limits the performance of supervised methods. There are two public datasets for screen image demoir\'{e}ing, which are TIP2018 dataset \cite{sun2018moire} and LCDMoire dataset \cite{yuan2019aim, yuan2019aimmethod}. TIP2018 is a real dataset with slight misalignment between each image pair and LCDMoire is a synthetic dataset. %, but there is a gap to the real dataset.
Because both datasets were developed for screen image demoir\'{e}ing, they are unsuitable to train a model to remove moir\'{e} patterns from images of high-frequency textures. 

To reduce moir\'{e} artifacts, some digital-camera manufacturers design special hardware, including special CFAs (Fuji's X-Trans, Sigma SD Quattro) and variable low-pass filters (Sony RX1RM2). 
Special CFAs and variable low-pass filters require special hardware design and thus cannot be widely used on smartphones.

In this paper, we propose a self-adaptive learning method for image demoir\'{e}ing. Our method removes moir\'{e} patterns from a moir\'{e} image with the help of a moir\'{e}-free blur image. We design a defocusing method to model the low-pass filter to obtain the blur image without moir\'{e} patterns.  
We use it as an additional input (defocused moir\'{e}-free image) to help remove moir\'{e} patterns from the moir\'{e} image (focused), and treat it as a joint filtering problem. 
% applicability
Our method can be easily applied to any digital camera. During the focusing process, a defocused blur image can be stored and combined with the focused image to perform image demoir\'{e}ing. Deep image prior \cite{lempitsky2018deep} shows that the structure of a generator network can capture a great deal of low-level image statistics without any training. In our model, we use a 3-layer fully connected sub-network to generate a blur kernel, and adopt a U-Net-like encoder/decoder architecture to perform image demoir\'{e}ing. Neither network is learned in an explicit training stage; but rather they are learned at test-time through an iterative, self-adaptive optimization.

In summary, our main contributions are:
\begin{enumerate}
\item We propose a self-adaptive learning method for image demoir\'{e}ing, which uses an additional input (defocused moir\'{e}-free image) to help remove moir\'{e} patterns from the focused moir\'{e} image.
\item We create a new dataset with pairs of focused moir\'{e} and defocused moir\'{e}-free images, containing both screen moir\'{e} images and high-frequency texture moir\'{e} images\footnote{In this paper, the terms `moir\'{e} image' and `moir\'{e}-free image' denote an image with and without moir\'{e} patterns. In addition, `screen moir\'{e} image' means an image whose moir\'{e} patterns is caused by digital screen, and `texture moir\'{e} image' denotes an image with moir\'{e} patterns caused by high-frequency textures.}.
\item Quantitative and qualitative experimental results on both public and our datasets show that our model outperforms state-of-the-art methods.
\item Our method, without a training stage, can be easily applied to any digital camera or smartphone.
\end{enumerate}	

\section{Related Work}
\label{gen_inst}
In this section, we review the most relevant work, including image demoir\'{e}ing, joint filtering, self-adaptive learning, and blind deblurring.

\textbf{Image Demoir\'{e}ing.} There are two common scenarios: \emph{screen image} demoir\'{e}ing and \emph{texture image} demoir\'{e}ing. 
Screen image demoir\'{e}ing focuses on removing moir\'{e} patterns from photos taken from screens, where moir\'{e} patterns are mainly caused by the interference between the screen's subpixel layout and the camera's color filter array.
Texture image demoir\'{e}ing deals with moir\'{e} patterns that are produced by photographing high-frequency scene content (e.g., fabric and long-distance buildings), which interferes with the CFA. 
Early work~\cite{sidorov2002suppression,sasada2003stationary,siddiqui2009hardware} on screen image demoir\'{e}ing focus on certain specific moir\'{e} patterns (striped, dotted or monotonous moir\'{e} patterns).
Recently, some deep learning models \cite{sun2018moire,liu2018demoir,he2019mop,liu2020wavelet,zheng2020image} cast screen demoir\'{e}ing as an image restoration problem and can handle more types of moir\'{e} patterns. 
Liu \textit{et al.}~\cite{liu2018demoir} built a coarse-to-fine convolutional neural network to remove moir\'{e} patterns from photos taken from screens.
Sun \textit{et al.}~\cite{sun2018moire} proposed a multi-resolution convolutional neural network for demoir\'{e}ing and released an associated dataset. 
He \textit{et al.}~\cite{he2019mop} labeled the data in \cite{sun2018moire} with three attribute labels of moir\'{e} patterns, which is beneficial to learn diverse patterns.
These methods are all supervised and need training with a large-scale dataset. Moreover, after training, they cannot generalize well to texture image demoir\'{e}ing. % 

Unlike screen image demoir\'{e}ing, removing moir\'{e} patterns in texture images is more challenging as moir\'{e} patterns appear only at the high-frequency areas and are always mixed with the underlying textures. 
Recently, some researchers \cite{yang2017textured,liu2015moire,yuan2020ntire} have attempted to handle texture image demoir\'{e}ing. 
Yang \textit{et al.}~\cite{yang2017textured} and Liu \textit{et al.}~\cite{liu2015moire} tried to remove moir\'{e} artifacts using low-rank and sparse matrix decomposition. 
Moir\'{e} patterns are also common artifacts from the image signal processing pipeline in a camera, especially from image demosaicing \cite{gharbi2016deep,liu2020joint}. Gharbi \textit{et al.}~\cite{gharbi2016deep} proposed to alleviate the moir\'{e} artifacts by fine-tuning their demosaicing model on a moir\'{e}-prone dataset, which was collected by measuring the frequency change from the ground-truth image and the demosaiced image.
Our network uses an additional input (defocused moir\'{e}-free image) to help remove moir\'{e} patterns from the moir\'{e} image and does not need training.

\textbf{Joint Filtering.}
Joint filtering has been applied to many low-level vision tasks \cite{petschnigg2004digital,gu2017learning}, with the aim of leveraging the guidance image as a prior and transferring the structural details to the target image. It has good ability in handling images from different domains. Many applications have been tried including depth/RGB image restoration \cite{gu2017learning}, flash/no-flash image denoising \cite{petschnigg2004digital}, texture removal \cite{ham2015robust,zhang2014rolling}, etc.
Local joint filtering methods \cite{tomasi1998bilateral,durand2002fast,he2013guided,zhang2014rolling} make use of a locally linear model to explore the relationship among neighboring pixels. Representative methods include bilateral filtering \cite{tomasi1998bilateral,durand2002fast} and guided filtering \cite{he2013guided}. But these methods usually introduce erroneous structures into the target image because they only explore the local structures of the guidance image.
Global joint filtering methods \cite{guo2018mutually,ferstl2013image,yan2013cross} optimize a global objective function. Different hand-crafted priors were proposed to enforce the target image and guidance image to have similar structures. But the hand-crafted priors may not reflect inherent structural details in the target image. 
Recently, some deep learning based joint filtering algorithms \cite{li2016deep,wu2018fast,pan2019spatially} have been proposed and shown better results. 
Pan \textit{et al.} \cite{pan2019spatially} presented spatially variant linear representation coefficients, which are determined by both the guidance image and the input image, to decide whether the structural details should be transferred to the output image.

Our method can also be viewed as a joint filtering method, taking two inputs (a focused moir\'{e} image and a defocused blur image). 
The defocused blur image provides important structural information to guide the demoir\'{e}ing network. The moir\'{e} patterns, especially low-frequency patterns, can be treated as a new structure overlaid on the structures of the moir\'{e}-free images. With the defocused image as a guide, the demoir\'{e}ing network can enhance the original structural information and suppress the moir\'{e} structures.
Unlike these deep learning based joint filtering methods, ours does not need training and produces a moir\'{e}-free image only from a defocused and focused image pair.

\textbf{Self-Adaptive Learning.}
Self-adaptive learning has been used in some specific low-level vision tasks such as super-resolution \cite{shocher2018zero,huang2015single,chengzero}, deblurring~\cite{michaeli2014blind,bahat2017non,ren2020neural}, inpainting~\cite{zhang2019internal} and dehazing~\cite{bahat2016blind}. 
They exploit the internal recurrence of information in an image without training. 
Recently, some researchers proposed some frameworks which can deal with multiple low-level vision tasks.
Lempitsky \textit{et al.}~\cite{lempitsky2018deep} showed that the structure of the deep image prior (DIP) neural network is very good to capture the low-level statistics of a single natural image. 
Gandelsman \textit{et al.}~\cite{gandelsman2019double-dip:} proposed a double-DIP framework for decomposing a single image into two layers. 
However, these two networks are unable to generate good results in the demoir\'{e}ing problem, partly because moir\'{e} patterns are widely distributed in the spatial and frequency domains.

\textbf{Blind Deblurring.}
Blind image deblurring is a very challenging
problem because it needs to estimate both the blur kernel and the clean image from a blur image. Blind image deblurring methods can be divided into optimization-based and deep-learning based.
Optimization-based methods use different priors for modeling clean images, such as gradient-based prior \cite{Pan2017L0,zuo2016learning}, patch-based prior \cite{michaeli2014blind,sun2013edge-based} and dark channel prior \cite{yan2017image,pan2018deblurring}.
For modeling accurate blur kernels, gradient sparsity prior \cite{levin2009understanding,pan2018deblurring} and spectral prior \cite{liu2014blind} are usually adopted.
In most cases, it is a blind deblurring problem to deblur a defocused image from an uncalibrated camera. 
Unlike these methods, we can obtain a clear image with moir\'{e} patterns through focusing. This image contains rich details and benefits restoring a clean moir\'{e}-free image. We also design a generative network to estimate the blur kernel.

\section{The Proposed Method}
\label{Method}
We first introduce the formulation of the problem, then design the network structure and finally present our algorithm.

\subsection{Problem Formulation}
When the image is defocused, image blur is spatially invariant in the same depth and the blur image $B$ can be formulated as, 
\begin{equation}
B =K \otimes C+N_{\sigma},
\label{blur}
\end{equation}
where $C$ is the underlying clean image, $K$ is the blur kernel, $N_{\sigma}$ is the additive Gaussian noise with noise level $\sigma$, and $\otimes$ denotes 2D convolution. In most cases where the camera is not calibrated, we need to estimate both $K$ and $C$ from a blur image $B$, which is an ill-posed problem.

When the image is focused, moir\'{e} artifacts may appear in high-frequency areas. To remove the moir\'{e} patterns, we suppose an image demoir\'{e}ing network $D$ can obtain the clean image such that
\begin{equation}
C=D\left(M\right),
\label{moire}
\end{equation}
where $M$ is a focused image contaminated with moir\'{e} patterns.
Putting Eqns. \ref{blur} and \ref{moire} together, we have
\begin{equation}
B =K \otimes D\left(M\right)+N_{\sigma}.
\label{combine}
\end{equation}
Thus, the blur image $B$ is obtained by first removing the moir\'{e} patterns from $M$ and then convolving with the blur kernel $K$. 
In this combined formulation, Eqn \ref{combine}, the demoir'{e}ing network $D$ can be learned without using the underlying clean image $C$ as the ground truth.

Inspired by the DIP framework~\cite{lempitsky2018deep}, we propose to use a generative network $G_{k}$ to capture the blur kernel $K$ as a prior. Finally the demoir\'{e}ing problem is formulated as  
\begin{equation}
\min _{D, G_{k}}\left\|G_{k}\left(\mathbf{z}\right) \otimes D\left(M\right)-B\right\|^{2},
\label{minimize}
\end{equation}
where $\mathbf{z}$ is a fixed vector and is sampled from the uniform distribution [0,1]. $G_{k}\left(\mathbf{z}\right)$ is the estimated blur kernel using the generator $G_{k}$. 
However, only optimizing Eqn. \ref{minimize} cannot guarantee that $D$ will give good demoir\'{e}ing results. Following \cite{levin2009understanding,pan2018deblurring}, we add the following constraints to the blur kernel,
\begin{equation}
\left(G_{k}\left(\mathbf{z}\right)\right)_{i} \geq 0, \ \forall i , 
\label{constraint1}
\end{equation}
\begin{equation}
 \sum_{i}\left(G_{k}\left(\mathbf{z}\right)\right)_{i}=1, 
\label{constraint2}
\end{equation}
where $(G_{k}(\cdot))_{i}$ denotes the $i$-th element in the blur kernel. 
Note that in Section \ref{Dataset} we will describe how to obtain $B$.

\begin{figure}[t]
  \centering
  \includegraphics[width=1.0\textwidth]{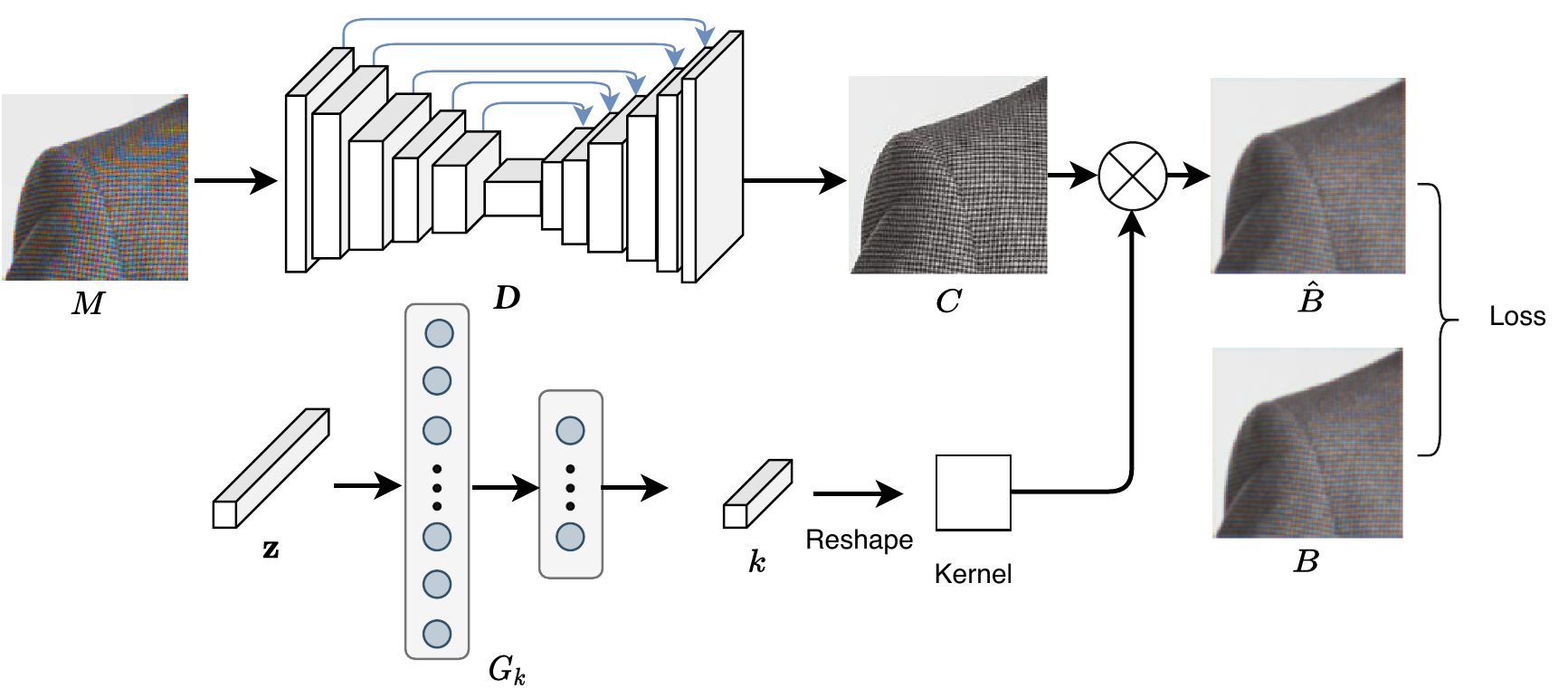}
  \caption{ Illustration of our model. The generative network $G_{k}$ is utilized to obtain a prior of the blur kernel. The network $D$ can output a moir\'{e}-free image after multiple iterations. The whole network is self-learned using only the focused image $M$ with moir\'{e} patterns and the defocused image $B$ without moir\'{e} patterns.}
  \label{fig:network}
\end{figure}

\subsection{Network Structure}

Fig. \ref{fig:network} shows the structure of our model. 
The input of $D$ is a focused image with moir\'{e} patterns. $D$ is a U-Net-like network where its first 5 layers of the encoder are connected via skip-connections to the 5 layers of the decoder. A convolutional output layer with the sigmoid function is used to generate the moir\'{e}-free image $C$. 
U-Net-like structures have been shown to work well in many low-level computer vision tasks \cite{isola2017image,sun2018moire,lempitsky2018deep}. 
For the network $G_{k}$, a blur kernel usually contains much less information than an image, and can be well estimated by a simpler generative network. 
Thus, we adopt a 3-layer fully-connected network (FCN) to serve as $G_k$.
It takes a 200-dimentional vector (noise) $\mathit{z}$ as the input.
The hidden layer and the output layer have 1,000 nodes and $K^{2}$ nodes, respectively, and the blur kernel size is $K \times K$. A softmax layer is applied to the output layer of $G_{k}$ to ensure the constraints in Eqns. \ref{constraint1} and \ref{constraint2}.

\subsection{Optimization Algorithm}
The optimization process of Eqn. \ref{minimize} can be viewed as a self-adaptive learning method. With only the defocused moir\'{e}-free image $B$ and the focused moir\'{e} image $M$, the networks $D$ and $G_{k}$ iteratively find better weights, making $D$ produce clear images $\hat{C}$ with fewer and fewer moir\'{e} patterns.
The parameters of $D$ and $G_{k}$ are simultaneously updated by back-propagation. Based on extensive experiments, we find that the joint optimization of $D$ and $G_{k}$ is better than the alternating optimization\footnote{The detail processes of the joint optimization and the alternating optimization and their difference are explained in the supplementary material.} of them (see the ablation study in Section \ref{sec:ablation}). 

\section{A New Dataset}
\label{Dataset}

We create a new dataset to evaluate our method quantitatively and qualitatively, as there is no public dataset available for the specific setting in this paper.

\textbf{Synthetic Data:} 
For the screen moir\'{e} patterns, the data are sampled from TIP2018 dataset \cite{sun2018moire}, from which we randomly choose 130 image pairs (moir\'{e} images and moir\'{e}-free images). The moir\'{e} images serve as the focused moir\'{e} images and the moir\'{e}-free images serve as the ground truth. 
To synthesize a defocused moir\'{e}-free image, we apply a Gaussian smoothing kernel (with $\sigma$ from 0.8 to 1.6) and an additive Gaussian noise (with noise level from 0 to 0.2) to the moir\'{e}-free image using Eqn. \ref{blur}. 
We assume a simplified case where the whole image has the same blurriness.
We call this subset \textit{SynScreenMoire}.

For texture moir\'{e} patterns, %as there is no publicly available dataset, 
we collect 30 high-quality images (with dense and regular textures) from the Internet and treat them as ground truth. 
The method in \cite{yang2017textured} is adopted to synthesize the corresponding moiré images. Finally, we use the same method for the screen moiré images to synthesize the defocused images from the ground truth. 
This subset is called \textit{SynTextureMoire}.

\begin{figure}[t]
  \centering
  \includegraphics[width=1.0\textwidth]{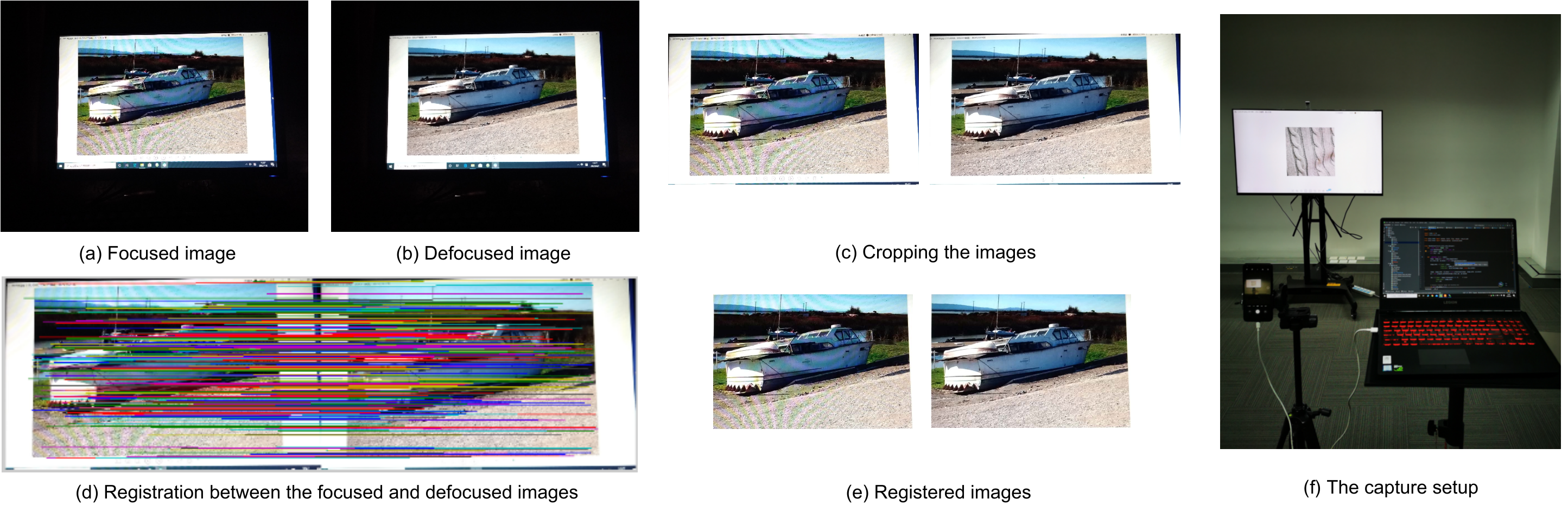}
  \caption{ Illustration of image acquisition.}
  \label{fig:acq}
\end{figure}

\textbf{Real Data:} 
We build a real dataset with 100 pairs, each with a focused moir\'{e} image and a defocused moir\'{e}-free image. 
It includes 50 pairs where the moir\'{e} patterns are caused by the interference between the camera CFAs and the screen pixel layouts (this subset is called \textit{RealScreenMoire}), and another 50 pairs where the moir\'{e} patterns are caused by the interference between the camera CFAs and the high-frequency textures of the images (this subset is called \textit{RealTextureMoire}). As shown in Fig. \ref{fig:acq}, to capture a pair of images, we design an image acquisition pipeline, which mainly consists of two steps: image capture and image alignment. %through focusing and defocusing, and (2) image.

\textbf{(1) Image Capture.} Each image is displayed at the centre of a computer screen (Fig. \ref{fig:acq}(a) and (b)) and the background color of the screen is set to white for better alignment. To produce a wide variety of moir\'{e} patterns, we use three types of smartphones (OPPO R9, HONOR 9 and HUAWEI P30 PRO). 
For different image pairs, we randomly change the distance and angle between the camera and the computer screen. The cameras are placed on a tripod. 
It is worth noting that when capturing texture moiré images, the camera is farther away from the screen than when acquiring screen moiré images, to avoid screen moiré patterns. By adjusting the distance between the camera and the screen, when the displayed high-frequency image textures (perceived by the camera) have frequencies similar to the camera's CFA, texture moir\'{e} patterns appear. Since the frequencies of the image textures are much lower than the frequencies of the computer screen's subpixel layout, screen moir\'{e} patterns are minimized.
In order to avoid camera shaking when changing the focus and defocus settings, we use a laptop to remotely control the zooming and shooting of the mobile phones. The capture process is shown in Fig. \ref{fig:acq}(f). We use a screen with 4k resolution to display the images when we make the \textit{RealTextureMoire} subset. % because we want to make the image display as clear as possible.
Using different phone or camera models as our capture devices ensures that the moir\'{e} patterns are across different optical sensors, while the diversity of display screens for \textit{RealScreenMoire} exhibits the difference in screen resolution\footnote{The detailed information about the camera models and the screens is described in the supplementary material.}. 
\textbf{(2) Image Alignment.}
%Images' scales will change when changing the focal length through focusing and defocusing.
As the focal length increases, the objects in the image will appear larger. We need to register each defocused and focused image pair.
With the help of a white background, we first binarize the captured image to find its area and then crop it.
%and obtain the magnification. We align image pairs using the magnification and foreground image boundaries.
We then align an image pair using the homography \cite{hartley2003multiple}.
% sift and ransac

\section{Experiments}
\label{experiment}

%To assess the abilities and quality of the proposed method, 
In this section, we show an ablation study and comparisons with state-of-the-art methods. 
%We conduct extensive ablation studies, and compare with state-of-the-art methods.
%we apply a series of experiments on both synthetic data sets and real data sets. We compare our method with state-of-the-art demoir\'{e}ing only methods and joint filtering methods. 
%
%\subsection{Implementation Details}
Our algorithm is implemented in Pytorch. The
experiments are conducted on a NVIDIA RTX 2080Ti GPU.
%
%The number of iterations does not need to be specific set according to different images and 
%We set the number of iterations to 3000.
%
%In the synthetic data set, we set the variance of the Gaussian blur kernel of blurred images to be between 0.8 and 1.6, and the noise variance to be 0.2.
%
In the kernel generation network, $\mathbf{z}$ is sampled from the uniform distribution in [0,1] with a fixed random seed 0. 
The initial learning rate is set to 0.01 and reduced by a half for every 500 iterations. The algorithm runs for 3000 iterations for each image pair.

\subsection{Ablation Study}
\label{sec:ablation}
\begin{figure}[t]
  \centering
  \includegraphics[width=1.0\textwidth]{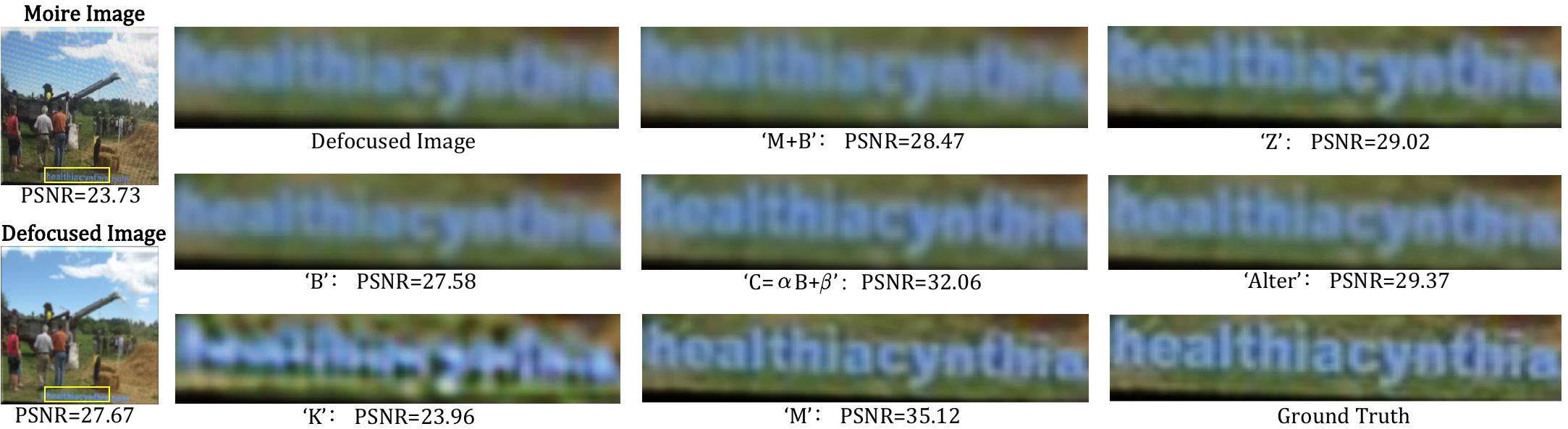}
  \caption{ One example of the ablation study.}
  \label{fig:abtest}
\end{figure}

\begin{table}[t]
  \centering\small
  
  \label{tab:abstudy2}
  \resizebox{13.0cm}{!} {
  \begin{tabular}{lccccccc}
    \toprule
      Network &M+B &Z&B&$\mathrm{C} = \alpha \mathrm{B}+\beta$&Alter & K &M \\                                       
     
     \midrule 
     PSNR/SSIM &27.43/0.852&28.42/0.869&27.17/0.842&29.57/0.846&27.31/0.855 &26.10/0.734 &$\mathbf{31.77/0.926}$\\
     
    \bottomrule
    
  \end{tabular}}
  \caption{Ablation study on \textit{SynScreenMoire}.}
  \label{tab:abstudy1}
\end{table}

The \textit{SynScreenMoire} subset is used to verify that the network $D$ can extract sufficient information from the focused image. % and that the information from the defocused image is redundant for $D$. 
%In order to verify that the network $D$ can extract sufficient recovery information from the focused image and the information of the defocused image is redundant for $D$, we conducted an ablation study on some images of the TIP2018 data set.
%
The results are shown in Table \ref{tab:abstudy1}, where `M+B' means that the input to the network is the concatenation of the focused and the defocused images; `B' denotes the input is only the defocused image; `Z' means the input is a 2D noise image (sampled from the uniform distribution in [0,1]) of the same size as the focused image; `M' stands for our original input (focused moir\'{e} image). 
Comparing `M' with `M+B' shows that adding the defocused moir\'{e}-free image as an additional input decreases the performance. 
%We speculate that the neural network cannot directly learn the same feature from different blur images. 
We speculate that the whole network is confused by this blur image in the input.
This also verifies the necessity of learning a blur kernel.
Comparing `M' with `Z' and `B' shows that the network $D$ does extract useful information from the focused moir\'{e} image.
Adding the moire image provides a good local minimum for the network to converge to. In fact, although the moir\'{e} image is corrupted by moire patterns, it still retains many high-frequency details. The U-NET can retain these details in the iteration for the reconstruction of a clean image from the blur image that lacks high-frequency details.
As shown in Fig. \ref{fig:abtest}, the result from `M' is better than those from `Z' and `B'. 
%
%In some tasks of low-level vision, predicting some parameters of images has advantage over predicting the final image directly \cite{wang2019underexposed,pan2019spatially}. 
Another baseline is by treating the final result as a spatially variant linear representation \cite{pan2019spatially} of the defocused moir\'{e}-free image ($
\mathrm{C} = \alpha \mathrm{B}+\beta$). %The results are shown in Table \ref{tab:abstudy1}. %,
The result shows that our end-to-end training (`M') is more powerful than predicting the parameters ($\alpha$ and $\beta$) of the image. 
%\textbf{Optimization Method} 
We also compare our joint optimization method with the alternating optimization method (`Alter' in Table \ref{tab:abstudy1})%\footnote{The details are shown in the supplementary material} %We evaluate the performance of our method using Algorithm 1 (joint optimization method) and alternating optimization method . 
, which optimizes $D$ and $G_k$ alternately (see the supplementary material for more details).  %demonstrating the superiority of our joint optimization method.
Moreover, we test the necessity of using a network (FCN) to generate the blur kernel by replacing the FCN with a learnable kernel, noted as `K' in Table \ref{tab:abstudy1}. The result shows that the PSNR/SSIM of `K' is smaller than that of `M'. Adding the FCN to learn the blur kernel can make the image satisfy the total variation prior and smooth the noise.
In addition, we also tried the other architecture of $D$, the encoder-decoder and ResNet. Their PSNRs are 0.11dB and 1.05dB, respectively lower than U-NET. Many other image restoration methods have shown that U-NET has advantages over the two structures.

\begin{table}[t]
  \centering\small
  \resizebox{14.0cm}{!}{
  \begin{tabular}{lcccccccccc}
    \toprule
      &Method  &   DMCNN~\cite{sun2018moire}  &  CFNet~\cite{liu2018demoir}  &  MopNet~\cite{he2019mop}   
      &  DIP~\cite{lempitsky2018deep}  & GF~\cite{he2013guided}& DJF~\cite{li2016deep} & MSJF~\cite{shen2015mutual-structure} & FDNet\\
      \midrule
      &S or U? & S &S &S &U &U &S &U &U\\
     \midrule
     \textit{SynScreenMoire} &PSNR/SSIM  & 26.15/0.869 &25.62/0.820 &	26.45/0.856 &22.57/0.757 & 27.23/0.808 &31.06/0.898&22.82/0.785&$\mathbf{31.77/0.926}$\\
    
     \midrule
     \textit{SynTextureMoire} &PSNR/SSIM  & 22.79/0.714 & 22.08/0.702 & 23.44/0.789 &22.51/0.720& 21.99/0.525&22.40/0.752 &24.70/0.687 &$\mathbf{25.98/0.794}$\\
    \bottomrule
  \end{tabular}}
  \caption{Quantitative comparison on \textit{SynScreenMoire} and \textit{SynTextureMoire}. S and U in the second row refer to Supervised and Unsupervised, respectively. The best results are highlighted in bold.}
  \label{tab:synresult1}
      
\end{table}

\begin{table}[t]
  \centering\small
  \resizebox{14.0cm}{!}{
  \begin{tabular}{lccccccccc}
    \toprule
         & Method   & MopNet~\cite{he2019mop}& DIP~\cite{lempitsky2018deep} &	DoubleDIP~\cite{gandelsman2019double-dip:} & GF~\cite{he2013guided}& DJF~\cite{li2016deep} & MSJF~\cite{shen2015mutual-structure} &SVLRM~\cite{pan2019spatially}& FDNet\\
      \midrule
      &S or U? &S &U  &U &U &S &U &S &U\\
     \midrule
     \textit{RealScreenMoire} &NIQE/BRISQUE  &5.57/30.53 &5.35/33.89 & 5.69/45.92& 6.42/45.91 &5.75/30.96&\underline{5.34}/\textbf{29.34}&9.42/31.04&$\mathbf{5.11}$/\underline{29.87}\\
     
     \textit{RealTextureMoire} &NIQE/BRISQUE  &17.85/42.81 &18.73/42.53 & 13.64/48.44&	11.99/51.84 &21.65/\underline{42.13}&\underline{11.67}/45.21&12.63/42.74&$\mathbf{11.22}/\mathbf{41.77}$\\
     
    \bottomrule
  \end{tabular}}
  \caption{Quantitative comparison of image demoir\'{e}ing on \textit{RealScreenMoire} and \textit{RealTextureMoire}. The best results are highlighted in bold and the second best are underlined.}
  \label{tab:realresult}
      
\end{table}

\subsection{Comparison with State-of-the-Art}
We compare our method with state-of-the-art demoir\'{e}ing methods (DMCNN~\cite{sun2018moire}, CFNet~\cite{liu2018demoir} and MopNet~\cite{he2019mop}), joint filtering methods (GF~\cite{he2013guided}, MSJF~\cite{shen2015mutual-structure}, DJF~\cite{li2016deep} and SVLRM~\cite{pan2019spatially}) and unsupervised image restoration methods (DIP~\cite{lempitsky2018deep} and DoubleDIP~\cite{gandelsman2019double-dip:}). %We use their source code for comparison. 
We also compare with some state-of-the-art blind deblurring methods to show that our model obtains useful information from the focused moir\'{e} images. For all blind deblurring methods and our method, the blur kernel and noise level are unknown.
In what follows, we term our model FDNet since it uses both the Focused and Defocused image pair for demoir\'{e}ing.

%\subsection{Quantitative Evaluation}
%\textbf{Full-reference evaluation.} 
\textbf{Evaluation on SynScreenMoire and SynTextureMoire.} On the synthetic data with ground truth, we can use the Peak Signal-to-Noise Ratio (PSNR) and the Structural Similarity Index Measure (SSIM) to compare the restored images.
%exploit full-reference image quality metrics to compare the restored images with the ground-truth. 
%
%We use both the Peak Signal-to-Noise Ratio (PSNR) and the Structural Similarity Index Measure (SSIM).
%
%The former measures the amount of signal lost w.r.t. a reference signal (e.g., an image), the latter compares two images' similarity in terms of visually structured elements.
%
%In Table 1, we conduct experiments synthetic dataset TIP2018. %The process of producing blurry image is described in Chapter 3. 
%
The supervised methods, DMCNN, CFNet, MopNet and DJF are trained on TIP2018 for testing on \textit{SynScreenMoire}, and trained on MITMoire for testing on \textit{SynTextureMoire}.
As shown in Table \ref{tab:synresult1}, FDNet outperforms all the state-of-the-art methods. % on image demoir\'{e}ing.
We also perform experiments on \textit{SynTextureMoire} to compare with the state-of-the-art blind deblurring methods to show FDNet extracts useful information from the focused moir\'{e} images. For DCP~\cite{pan2018deblurring} and DeblurGANv2~\cite{kupyn2019deblurgan-v2:}, the PSNRs/SSIMs are 28.53/0.875 and 28.58/0.864, respectively, which are smaller than FDNet's 31.77/0.926. 

\textbf{Evaluation on RealScreenMoire and RealTextureMoire.} On the real data without ground truth, we evaluate all generated images using no-reference quality metrics, which estimate absolute image quality scores. %, not a proximity to a reference. 
For objective quality measurement, we use the Blind/Referenceless Image Spatial Quality Evaluator (BRISQUE)~\cite{mittal2011blindreferenceless} and Naturalness Image Quality Evaluator (NIQE)~\cite{mittal2012making}. BRISQUE extracts the point-wise statistics of local normalized luminance signals and measures image naturalness. NIQE is based on the construction of a quality-aware collection of statistical features based on a simple space domain natural scene statistic model.
%
%To evaluate the clarity of the image (we do not want to get blurry image), we also use image Variance to evaluate image quality.
%
Note that a smaller NIQE or BRISQUE score means an image with better quality (lower is better). An off-the-shelf trained algorithm (MATLAB2018b) is used to obtain the NIQE and BRISQUE scores. 
To compare with the existing methods, we randomly choose 25 images and 38 images from our real dataset \textit{RealScreenMoire} and \textit{RealTextureMoire}, respectively. 
%To compare with supervised methods (DJF and SVLRM), we train them in TIP2018 or MITMoire first and test them on real data. 
The supervised methods MopNet, DJF and SVLRM are trained on TIP2018 for testing on \textit{RealScreenMoire}, and trained on MITMoire for testing on \textit{RealTextureMoire}.
As shown in Table \ref{tab:realresult}, our method overall outperforms both the supervised and unsupervised methods.
%
%It shows that our predicted images are not only closer to the natural images, but also clearer than those predicted by other methods.

\begin{figure}[t!]
  \centering
  \includegraphics[width=1.0\textwidth]{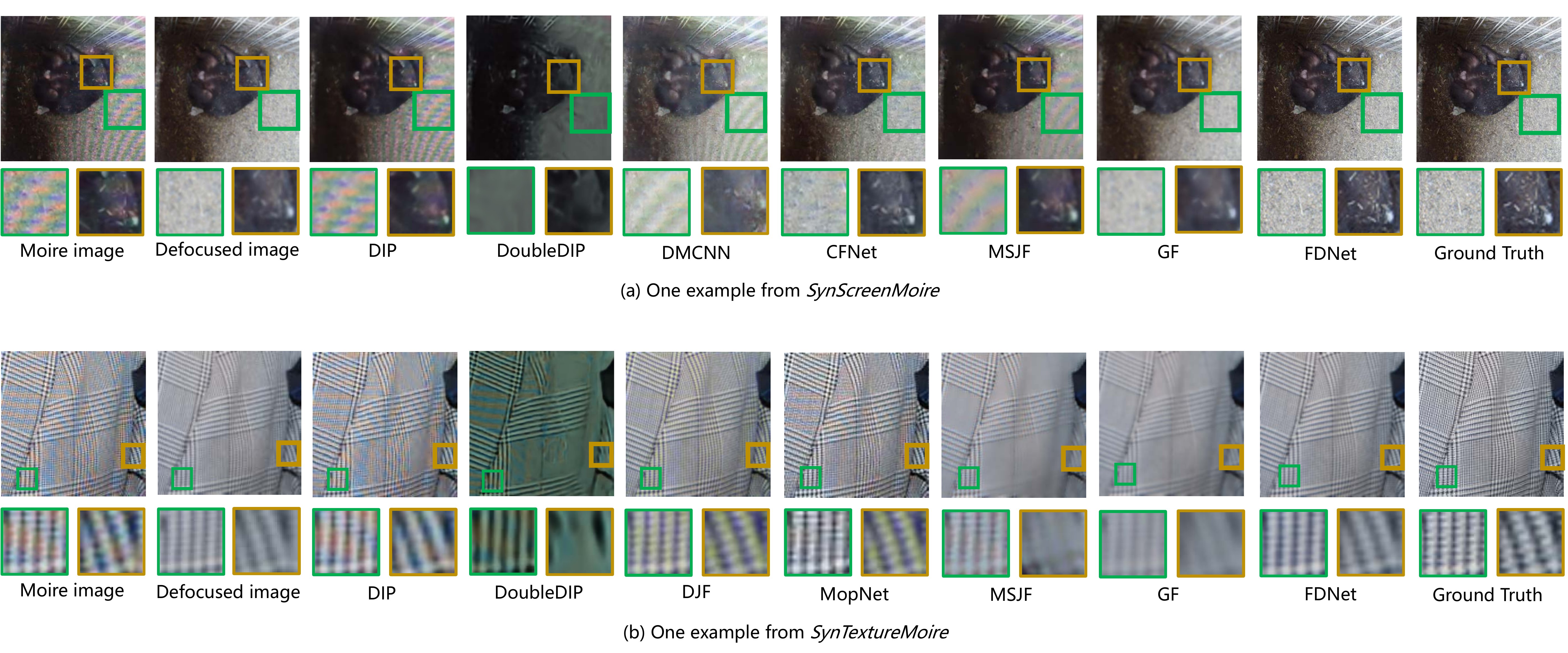}
  \caption{ Visual comparison among our FDNet and other models, evaluated on images from \textit{SynScreenMoire} and \textit{SynTextureMoire}.}
  \label{fig:imgresult1}
\end{figure}
\textbf{Qualitative Results.}  
As shown in Fig. \ref{fig:imgresult1}, the result of DIP has obvious moir\'{e} artifacts left, and DoubleDIP has a global color shift from the original input and over-smoothed details. 
%DoubleDIP almost removes the color of the image.
%
These two deep image prior methods cannot effectively remove moir\'{e} patterns, perhaps because they struggle to learn the low-frequency characteristics and the color diversity of moir\'{e} patterns. Moire patterns and noise are different; the former are prevalent more in low and mid-frequencies. DIP relies on the spectral bias of the CNN to learn lower frequencies first. So before DIP learns the high-frequency details of the image, moire patterns have appeared in the results of DIP.
%
%Some demoir\'{e}ing only methods (DMCNN, CFNet and MopNet) cannot effectively remove the moiŕe.
The demoir\'{e}ing only methods (DMCNN, CFNet and MopNet) cannot effectively remove the moir\'{e} patterns.
%produce better results on the synthenic test images, they make the real test images blurry.
% liulin: Enough space to show some real test data ???
%
In addition, the joint filtering methods (GF and MSJF) tend to blur the high-frequency regions and cannot remove the moir\'{e} patterns well with the guidance of the blur image. 
In contrast, our method FDNet eliminates the moir\'{e} patterns more effectively, benefiting from the accurate prediction of the blur kernel. 
In addition, FDNet retains the original textures in the images with moir\'{e} patterns
removed instead of over-smoothing the high-frequency regions. More results are provided in the supplementary material.
%\footnote{More results are provided in the supplementay material.}.

\subsection{Practical Applications of our Method}

Our method has the potential of being applied to smartphones without modifying the hardware. In the typical capture mode, the camera usually is embedded with an auto-focus algorithm, which can be modified to save an additional defocused image. 
Unlike variable hardware low-pass filters in some DSLR cameras that require user control, our method is invisible to the user.
The defocused image and the focused image can be used to perform image demoir\'{e}ing.
%[Shanxin: it will be nice to show this on real phone, e.g. P30, but it could be difficult to get the data? the defocused image during the auto-focus process.]

\section{Conclusion}
We have proposed a self-adaptive learning method for moir\'{e} pattern removal. 
Our network predicts a moir\'{e}-free clear image from a focused image with moir\'{e} patterns, with the help of a corresponding defocused moir\'{e}-free blur image, 
It substantially outperforms state-of-the-art demoir\'{e}ing methods and joint filtering methods.
The moir\'{e}-free blur image is easy to obtain through software or hardware. In addition, we have built the first dataset with pairs of focused moir\'{e} images and defocused moir\'{e}-free images.
The future work includes finding more accurate blur kernel estimation and more efficient restoration.

% %%%%%%%%  Original version
% %In order to provide a balanced perspective, authors are required to include a statement of the potential broader impact of their work, including its ethical aspects and future societal consequences. Authors should take care to discuss both positive and negative outcomes.
% \section{Broader Impact}
% %this section is allowed to go to page 9
% Our method can help digital cameras and smartphones to reduce moir\'{e} artifacts when taking photos from screens or from high-frequency natural scene content. 
% As a self-adaptive learning method, it can be easily embedded into digital cameras and smartphones without special hardware requirement. Using existing auto-focus algorithms, a defocused blur image (during the auto-focus) and the focused image (after the auto-focus) can be stored. 
% %
% In the case of hand-held image capturing, hand shaking and focal length change may lead to slight motion misalignment between the image pair, and then a global alignment (e.g., homography) may be needed. Additional computation is also required for running our algorithm on the image pair. 
% %
% %Applying our method in the raw domain, ..., TODO
% %%%negative outcomes
% %Our method aims at restoring the real content contaminated by moir\'{e} patterns. In an extreme case, when taking photos of printed moir\'{e} patterns with the aim of sepcifically saving/storing the moir\'{e} patterns, our method may remove them. (don't know...)
% % 

%%%%%%%%  Original version

\section*{Broader Impact}
%this section is allowed to go to page 9
Our method improves the quality of photographs taken from a digital camera, by removing moir\'{e} patterns to restore an underlying clean, moir\'{e}-free image.  By design, the algorithm produces restored images that are more faithful to the true scene. This makes the photograph's information more apparent and representative. It is envisioned better quality images will have a positive societal benefit, making visually recorded information more detailed, informative, and useful.  

With any approach that improves image quality also comes the risk of negative uses, such as privacy issues.  For images of natural scenes, further downstream applications such as surveillance and tracking may become more effective particularly in high-frequency regions of an image where moir\'{e} patterns are more likely.

Another consideration is that by removing moir\'{e} patterns from pictures taken of digital displays, it may become more difficult to determine, from the image alone, if it is taken of natural scene or of a display such as a computer screen.  Potentially this could make it easier for one to take a photograph of a digital screen and claim that the photo is an authentic capture of real scene.  However, there may be other indicators if the photo is taken of the screen, particularly if the layout of the LCD elements are visible.  Potential future research could explore the difficulty of classifying screen and natural images, with and without demoir\'{e}d results produced by the proposed method.

We note, although our method substantially improves the state-of-the-art, it is not perfect and its failures may result in moir\'{e} patterns to remain in an image, or be replaced with blurry outputs. As the method is self-adaptive, learning at test-time from two images, we believe the implications of learning from biased data to be minimal.

%As our method is self-adaptive, at inference, it is best suited a computing platform that has a GPU. The method requires two images to be taken, along with iterative solving at test-time.  This may have implications on memory, battery usage and heat if implemented directly on a smartphone. Alternative deployments include running the method on the cloud or on a PC after photos have been taken.

\section*{Funding Disclosure}
The work was supported by the Computer Vision Research Project of Huawei Noah's Ark Lab.

{\small
\bibliographystyle{plain}
\bibliography{egbib}

\begin{thebibliography}{10}

\bibitem{bahat2017non}
Yuval Bahat, Netalee Efrat, and Michal Irani.
\newblock Non-uniform blind deblurring by reblurring.
\newblock In {\em ICCV, 2017}.

\bibitem{bahat2016blind}
Yuval Bahat and Michal Irani.
\newblock Blind dehazing using internal patch recurrence.
\newblock In {\em ICCP, 2016}.

\bibitem{chengzero}
Xi~Cheng, Zhenyong Fu, and Jian Yang.
\newblock Zero-shot image super-resolution with depth guided internal
  degradation learning.

\bibitem{durand2002fast}
Fredo Durand and Julie Dorsey.
\newblock Fast bilateral filtering for the display of high-dynamic-range
  images.
\newblock {\em SIGGRAPH, 2002}.

\bibitem{ferstl2013image}
David Ferstl, Christian Reinbacher, Rene Ranftl, Matthias Ruther, and Horst
  Bischof.
\newblock Image guided depth upsampling using anisotropic total generalized
  variation.
\newblock In {\em ICCV, 2013}.

\bibitem{gandelsman2019double-dip:}
Yossi Gandelsman, Assaf Shocher, and Michal Irani.
\newblock “double-dip”: Unsupervised image decomposition via coupled
  deep-image-priors.
\newblock {\em CVPR, 2019}.

\bibitem{gharbi2016deep}
Micha{\"e}l Gharbi, Gaurav Chaurasia, Sylvain Paris, and Fr\'{e}do Durand.
\newblock Deep joint demosaicking and denoising.
\newblock {\em TOG, 2016}.

\bibitem{gu2017learning}
Shuhang Gu, Wangmeng Zuo, Shi Guo, Yunjin Chen, Chongyu Chen, and Lei Zhang.
\newblock Learning dynamic guidance for depth image enhancement.
\newblock In {\em CVPR, 2017}.

\bibitem{guo2018mutually}
Xiaojie Guo, Yu~Li, Jiayi Ma, and Haibin Ling.
\newblock Mutually guided image filtering.
\newblock {\em TPAMI, 2018}.

\bibitem{ham2015robust}
Bumsub Ham, Minsu Cho, and Jean Ponce.
\newblock Robust image filtering using joint static and dynamic guidance.
\newblock In {\em CVPR, 2015}.

\bibitem{hartley2003multiple}
Richard Hartley and Andrew Zisserman.
\newblock {\em Multiple view geometry in computer vision}.
\newblock Cambridge University Press, 2003.

\bibitem{he2019mop}
Bin He, Ce~Wang, Boxin Shi, and Ling-Yu Duan.
\newblock Mop moire patterns using mopnet.
\newblock In {\em ICCV, 2019}.

\bibitem{he2013guided}
Kaiming He, Jian Sun, and Xiaoou Tang.
\newblock Guided image filtering.
\newblock {\em TPAMI}, 2013.

\bibitem{huang2015single}
Jia-Bin Huang, Abhishek Singh, and Narendra Ahuja.
\newblock Single image super-resolution from transformed self-exemplars.
\newblock In {\em CVPR, 2015}.

\bibitem{isola2017image}
Phillip Isola, Jun-Yan Zhu, Tinghui Zhou, and Alexei~A Efros.
\newblock Image-to-image translation with conditional adversarial networks.
\newblock In {\em CVPR, 2017}.

\bibitem{kupyn2019deblurgan-v2:}
Orest Kupyn, Tetiana Martyniuk, Junru Wu, and Zhangyang Wang.
\newblock Deblurgan-v2: Deblurring (orders-of-magnitude) faster and better.
\newblock {\em ICCV, 2019}.

\bibitem{lempitsky2018deep}
Victor Lempitsky, Andrea Vedaldi, and Dmitry Ulyanov.
\newblock Deep image prior.
\newblock {\em CVPR, 2018}.

\bibitem{levin2009understanding}
Anat Levin, Yair Weiss, Fredo Durand, and William~T Freeman.
\newblock Understanding and evaluating blind deconvolution algorithms.
\newblock In {\em CVPR, 2009}.

\bibitem{li2016deep}
Yijun Li, Jiabin Huang, Narendra Ahuja, and Minghsuan Yang.
\newblock Deep joint image filtering.
\newblock ECCV, 2016.

\bibitem{liu2018demoir}
Bolin Liu, Xiao Shu, and Xiaolin Wu.
\newblock Demoir{\'e}ing of camera-captured screen images using deep
  convolutional neural network.
\newblock {\em arXiv preprint arXiv:1804.03809, 2018}.

\bibitem{liu2015moire}
Fanglei Liu, Jingyu Yang, and Huanjing Yue.
\newblock Moir{\'e} pattern removal from texture images via low-rank and sparse
  matrix decomposition.
\newblock In {\em VCIP, 2015}.

\bibitem{liu2014blind}
Guangcan Liu, Shiyu Chang, and Yi~Ma.
\newblock Blind image deblurring using spectral properties of convolution
  operators.
\newblock {\em TIP, 2014}.

\bibitem{liu2020joint}
Lin Liu, Xu~Jia, Jianzhuang Liu, and Qi~Tian.
\newblock Joint demosaicing and denoising with self guidance.
\newblock In {\em CVPR, 2020}.

\bibitem{liu2020wavelet}
Lin Liu, Jianzhuang Liu, Shanxin Yuan, Gregory Slabaugh, Ales Leonardis,
  Wengang Zhou, and Qi~Tian.
\newblock Wavelet-based dual-branch network for image demoir{\'e}ing.
\newblock In {\em ECCV, 2020}.

\bibitem{michaeli2014blind}
Tomer Michaeli and Michal Irani.
\newblock Blind deblurring using internal patch recurrence.
\newblock In {\em ECCV, 2014}.

\bibitem{mittal2011blindreferenceless}
Anish Mittal, Anush~K Moorthy, and Alan~C Bovik.
\newblock Blind/referenceless image spatial quality evaluator.
\newblock {\em ASILOMAR, 2011}.

\bibitem{mittal2012making}
Anish Mittal, Rajiv Soundararajan, and Alan~C Bovik.
\newblock Making a “completely blind” image quality analyzer.
\newblock {\em IEEE Signal Processing Letters, 2012}.

\bibitem{pan2019spatially}
Jinshan Pan, Jiangxin Dong, Jimmy Ren, Liang Lin, Jinhui Tang, and Minghsuan
  Yang.
\newblock Spatially variant linear representation models for joint filtering.
\newblock CVPR, 2019.

\bibitem{pan2018deblurring}
Jinshan Pan, Deqing Sun, Hanspeter Pfister, and Minghsuan Yang.
\newblock Deblurring images via dark channel prior.
\newblock {\em TPAMI, 2018}.

\bibitem{Pan2017L0}
Jinshan Pan, Hu~Zhe, Zhixun Su, and Ming~Hsuan Yang.
\newblock L0 -regularized intensity and gradient prior for deblurring text
  images and beyond.
\newblock {\em TPAMI, 2017}.

\bibitem{petschnigg2004digital}
Georg Petschnigg, Richard Szeliski, Maneesh Agrawala, Michael Cohen, Hugues
  Hoppe, and Kentaro Toyama.
\newblock Digital photography with flash and no-flash image pairs.
\newblock {\em TOG, 2004}.

\bibitem{ren2020neural}
Dongwei Ren, Kai Zhang, Qilong Wang, Qinghua Hu, and Wangmeng Zuo.
\newblock Neural blind deconvolution using deep priors.
\newblock In {\em CVPR, 2020}.

\bibitem{sasada2003stationary}
Ryoji Sasada, Masahiko Yamada, Shoji Hara, Hideya Takeo, and Kazuo Shimura.
\newblock Stationary grid pattern removal using 2d technique for moir{\'e}-free
  radiographic image display.
\newblock In {\em Medical Imaging, 2003}.

\bibitem{shen2015mutual-structure}
Xiaoyong Shen, Chao Zhou, Li~Xu, and Jiaya Jia.
\newblock Mutual-structure for joint filtering.
\newblock {\em ICCV, 2015}.

\bibitem{shocher2018zero}
Assaf Shocher, Nadav Cohen, and Michal Irani.
\newblock “zero-shot” super-resolution using deep internal learning.
\newblock In {\em CVPR, 2018}.

\bibitem{siddiqui2009hardware}
Hasib Siddiqui, Mireille Boutin, and Charles~A Bouman.
\newblock Hardware-friendly descreening.
\newblock {\em TIP, 2009}.

\bibitem{sidorov2002suppression}
Denis~N Sidorov and Anil~Christopher Kokaram.
\newblock Suppression of moire patterns via spectral analysis.
\newblock In {\em VCIP, 2002}.

\bibitem{sun2013edge-based}
Libin Sun, Sunghyun Cho, Jue Wang, and James Hays.
\newblock Edge-based blur kernel estimation using patch priors.
\newblock {\em ICCP, 2013}.

\bibitem{sun2018moire}
Yujing Sun, Yizhou Yu, and Wenping Wang.
\newblock Moir{\'e} photo restoration using multiresolution convolutional
  neural networks.
\newblock {\em TIP, 2018}.

\bibitem{tomasi1998bilateral}
Carlo Tomasi and Roberto Manduchi.
\newblock Bilateral filtering for gray and color images.
\newblock {\em ICCV, 1998}.

\bibitem{wu2018fast}
Huikai Wu, Shuai Zheng, Junge Zhang, and Kaiqi Huang.
\newblock Fast end-to-end trainable guided filter.
\newblock {\em CVPR, 2018}.

\bibitem{yan2013cross}
Qiong Yan, Xiaoyong Shen, Li~Xu, Shaojie Zhuo, Xiaopeng Zhang, Liang Shen, and
  Jiaya Jia.
\newblock Cross-field joint image restoration via scale map.
\newblock In {\em ICCV, 2013}.

\bibitem{yan2017image}
Yanyang Yan, Wenqi Ren, Yuanfang Guo, Rui Wang, and Xiaochun Cao.
\newblock Image deblurring via extreme channels prior.
\newblock {\em CVPR, 2017}.

\bibitem{yang2017textured}
Jingyu Yang, Fanglei Liu, Huanjing Yue, Xiaomei Fu, Chunping Hou, and Feng Wu.
\newblock Textured image demoir{\'e}ing via signal decomposition and guided
  filtering.
\newblock {\em TIP, 2017}.

\bibitem{yuan2020ntire}
Shanxin Yuan, Radu Timofte, Ales Leonardis, Gregory Slabaugh, et~al.
\newblock Ntire 2020 challenge on image demoireing: Methods and results.
\newblock In {\em CVPRW, 2020}.

\bibitem{yuan2019aim}
Shanxin Yuan, Radu Timofte, Gregory Slabaugh, and Ales Leonardis.
\newblock Aim 2019 challenge on image demoireing: Dataset and study.
\newblock {\em ICCVW, 2019}.

\bibitem{yuan2019aimmethod}
Shanxin Yuan, Radu Timofte, Gregory Slabaugh, Ales Leonardis, Bolun Zheng, Xin
  Ye, Xiang Tian, Yaowu Chen, Xi~Cheng, Zhenyong Fu, et~al.
\newblock Aim 2019 challenge on image demoireing: Methods and results.
\newblock In {\em ICCVW, 2019}.

\bibitem{zhang2019internal}
Haotian Zhang, Long Mai, Ning Xu, Zhaowen Wang, John Collomosse, and Hailin
  Jin.
\newblock An internal learning approach to video inpainting.
\newblock In {\em ICCV, 2019}.

\bibitem{zhang2014rolling}
Qi~Zhang, Xiaoyong Shen, Li~Xu, and Jiaya Jia.
\newblock Rolling guidance filter.
\newblock In {\em ECCV, 2014}.

\bibitem{zheng2020image}
Bolun Zheng, Shanxin Yuan, Gregory Slabaugh, and Ales Leonardis.
\newblock Image demoireing with learnable bandpass filters.
\newblock In {\em CVPR, 2020}.

\bibitem{zuo2016learning}
Wangmeng Zuo, Dongwei Ren, David Zhang, Shuhang Gu, and Lei Zhang.
\newblock Learning iteration-wise generalized shrinkage–thresholding
  operators for blind deconvolution.
\newblock {\em TIP, 2016}.

\end{thebibliography}
}

\appendix

\newpage

\section{Cameras and Screens}

We use three cameras and three screens to capture our dataset; please see Table~\ref{tab:diviceinfo} for the specifications. Note that the HONOR Intelligence Screen is a screen with 4K resolution and is used to display images for the \textit{RealTextureMoire} subset. A high-resolution screen helps avoid screen moiré patterns when acquiring texture moiré images. 
\begin{table}[h]
  \centering\small
  \caption{Camera specifications and screen specifications.}
  \label{tab:diviceinfo}
  \resizebox{13.0cm}{!} {
  \begin{tabular}{cccccc}
    \toprule
       \multicolumn{3}{c}{Capture device} & \multicolumn{3}{|c}{Display device} \\  
       
      Manufacturer &Model&Image Resolution&\multicolumn{1}{|c}{Manufacturer} &Model&Resolution\\
     \midrule 
      OPPO & R9 &$4608\times3456$&\multicolumn{1}{|c}{SAMSUNG}&S22F350H&$1920\times1080$\\
      HONOR & 9 &$3264\times1632$&\multicolumn{1}{|c}{HONOR}& Intelligence Screen &4K\\
      HUAWEI & P30 PRO &$3648\times2736$&\multicolumn{1}{|c}{HP}& E243 &$1920\times1200$\\
    \bottomrule
  \end{tabular}}
\end{table}

\section{The Alternating Optimization Method}
\label{Alternating}

The following algorithms (Procedure \ref{alg:Framework1} and Procedure \ref{alg:Framework2}) show the joint optimization method and the baseline alternating optimization method compared in the ablation study in Section 5.1 of the main paper. The difference between the joint optimization used for our FDNet and the alternating optimization is shown on the lines 6--10 of Procedure \ref{alg:Framework2}, where $D^{i}$ and $G_{k}^{i}$ are updated in an alternating fashion.

 \begin{algorithm}[H]  
  \caption{The joint optimization algorithm.}  
  \label{alg:Framework1}  
  \begin{algorithmic}[1]  
    \REQUIRE   
    Focused image $M$ with moir\'{e} patterns and defocused blur image $B$ without moir\'{e} patterns;  
    \ENSURE  
    Estimated moir\'{e}-free image $C$;  
    \STATE Initialize $G^{0}_{k}$ and $D^{0}$ with Gaussian random weights;
    \STATE Sample $\mathbf{z}$ from the uniform distribution [0,1];
    %\STATE $k = G^{0}_{k}(\mathit{z})$;
           \FOR{ $i$ = 1 to $N$}
    \STATE $\hat{C} = D^{i-1}(M)$; $\hat{k} = G^{i-1}_{k}(\mathbf{z})$; $\hat{B}=\hat{C} \otimes \hat{k}$ ;
    \STATE $Loss = MSE(\hat{B},B)$;
    \STATE Update $D^{i}$ and $G_{k}^{i}$ simultaneously using the ADAM algorithm;
           \ENDFOR
    \RETURN  $C = D^{N}(M)$.
  \end{algorithmic}  
\end{algorithm}   

 \begin{algorithm}[H]  
  \caption{The alternating optimization algorithm.}  
  \label{alg:Framework2}  
  \begin{algorithmic}[1]  
    \REQUIRE   
    Focused image $M$ with moir\'{e} patterns and defocused blur image $B$ without moir\'{e} patterns;  
    \ENSURE  
    Estimated moir\'{e}-free image $C$;  
    \STATE Initialize $G^{0}_{k}$ and $D^{0}$ with Gaussian random weights;
    \STATE Sample $\mathbf{z}$ from the uniform distribution [0,1];
    %\STATE $k = G^{0}_{k}(\mathit{z})$;
           \FOR{ $i$ = 1 to $N$}
    \STATE $\hat{C} = D^{i-1}(M)$; $\hat{k} = G^{i-1}_{k}(\mathbf{z})$; $\hat{B}=\hat{C} \otimes \hat{k}$ ;
    \STATE $Loss = MSE(\hat{B},B)$;
    \IF { $i$ is even} 
     \STATE       Update $D^{i}$ using the ADAM algorithm;
      $G_{k}^{i}=G_{k}^{i-1}$;
            \ELSE \STATE Update $G_{k}^{i}$ using the ADAM algorithm;
            $D^{i}=D^{i-1}$;
             \ENDIF
           \ENDFOR
    \RETURN  $C = D^{N}(M)$.
  \end{algorithmic}  
\end{algorithm}

\newpage
\section{Results from Real Natural Scenes}

\begin{figure}[H]
  \centering
  \includegraphics[width=1.0\textwidth]{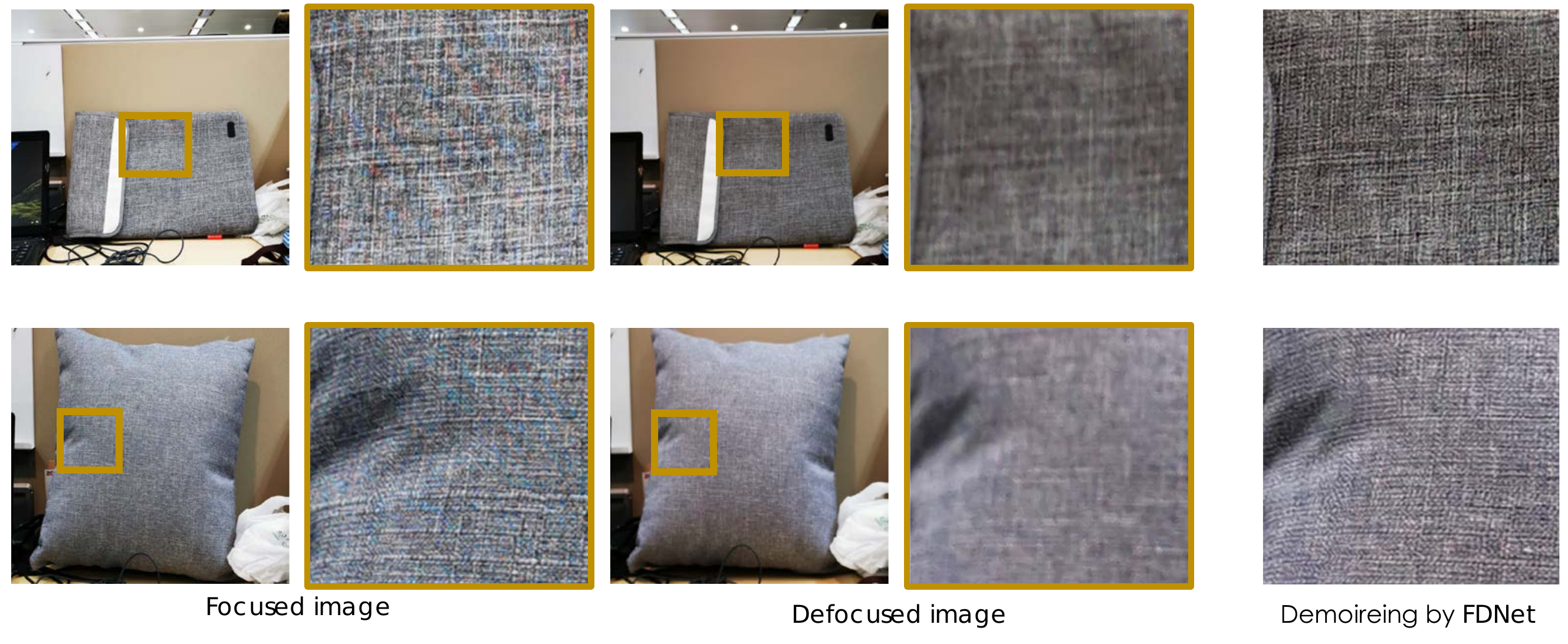}
  \includegraphics[width=1.0\textwidth]{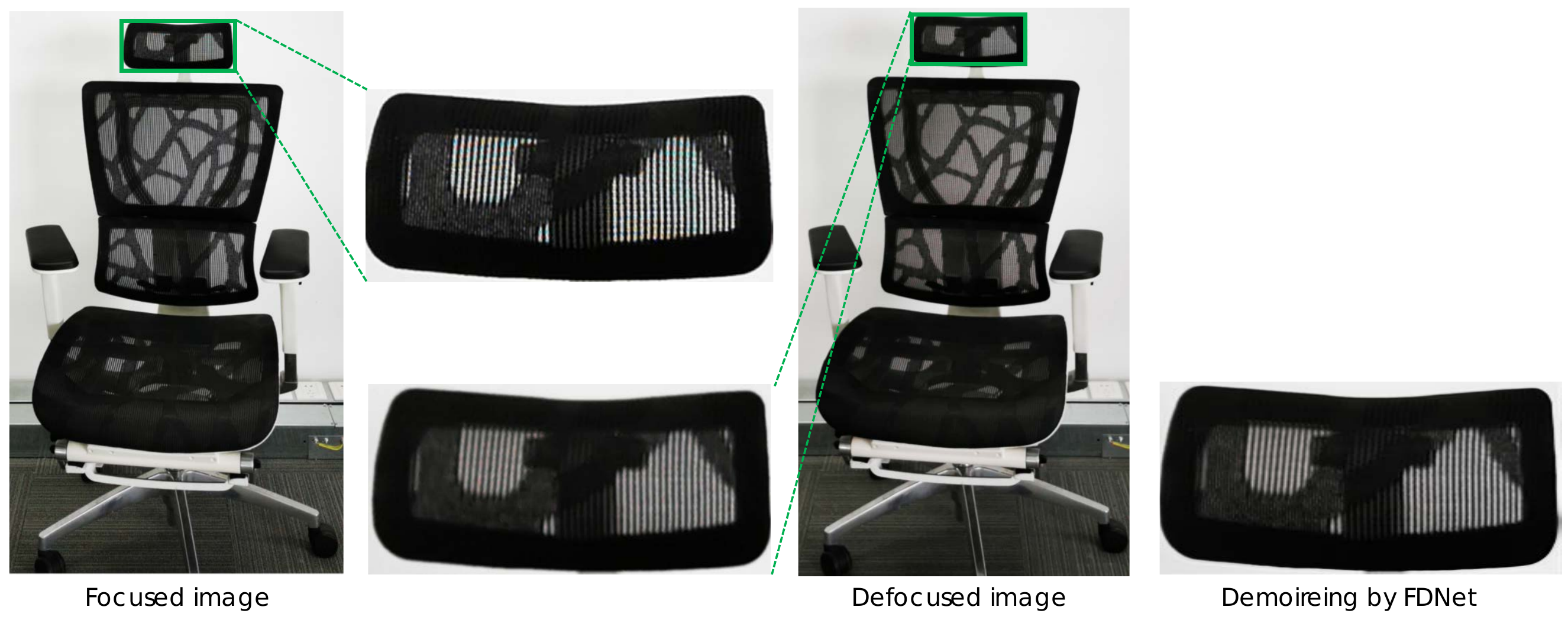}
  \caption{ Examples captured from natural scenes.}
  \label{fig:r2}
\end{figure}

We also test our model on a smartphone HUAWEI P30 PRO. We collect some focused and defocused image pairs from natural scenes, where the focused images have texture moir\'{e} patterns, as shown in  Figure \ref{fig:r2}. To test on the real world examples, we do some preprocessing, e.g., alignment. We keep the areas where the moire is produced at the same depth. FDNet generalizes well to images taken from natural scenes (not from screens), as the results are moir\'{e}-free and the details are retained from the focused moir\'{e} image.

\iffalse  % comment
\begin{figure}[H]
  \centering
  \includegraphics[width=1.0\textwidth]{realscene3.pdf}
  \caption{Focused and defocused image pairs of an office chair.}
  \label{fig:r1}
\end{figure}
\fi

\section{Efficiency Comparison among our Model, DIP and Double-DIP}
We evaluate the efficiencies of the FDNet and the deep-image-prior methods (DIP and DoubleDIP) on an NVIDIA RTX 2080Ti GPU. 
The number of iterations for DIP to find an optimal result varies from image to image, and needs to be manually adjusted. In the demoir\'{e}ing task, DIP takes 1000 iterations. 
Double-DIP also requires 1000 iterations to converge.
Our model does not have the problem of getting worse results when iterations are over some threshold, due to the constraint by the blur image.
FDNet converges in about 500 iterations and then its PSNR slightly increases with the iteration number increasing (see Figure \ref{fig:iter} for one example). 
As shown in Table \ref{tab:eff}, our FDNet has a faster runtime.
%As shown in Table \ref{tab:eff}, our FDNet is relatively small and fast. % ours is the fastest among all the 3 deep learning methods.

\begin{table}[H]
  \centering\small
  
  \begin{tabular}{lccc }
    \toprule
      Algorithm &  DoubleDIP  &   DIP      & FDNet    \\
     
     \midrule

     %Time\ (s/image)  &  280    & 43 & 174 \\
     %Time\ (s/iteration)  &  280    & 43 & 60 \\
     Time\ (s)  &  280    & 43 & 30 \\
     Parameters\ (MB)  &  3.08&	 2.22  &  2.64  \\
     
    \bottomrule
  \end{tabular}
     \caption{Efficiency comparison.}
  \label{tab:eff}  
\end{table}

\section{Visualization of Intermediate Results and Blur Kernels}
\begin{figure}[H]
  \centering
  \includegraphics[width=1.0\textwidth]{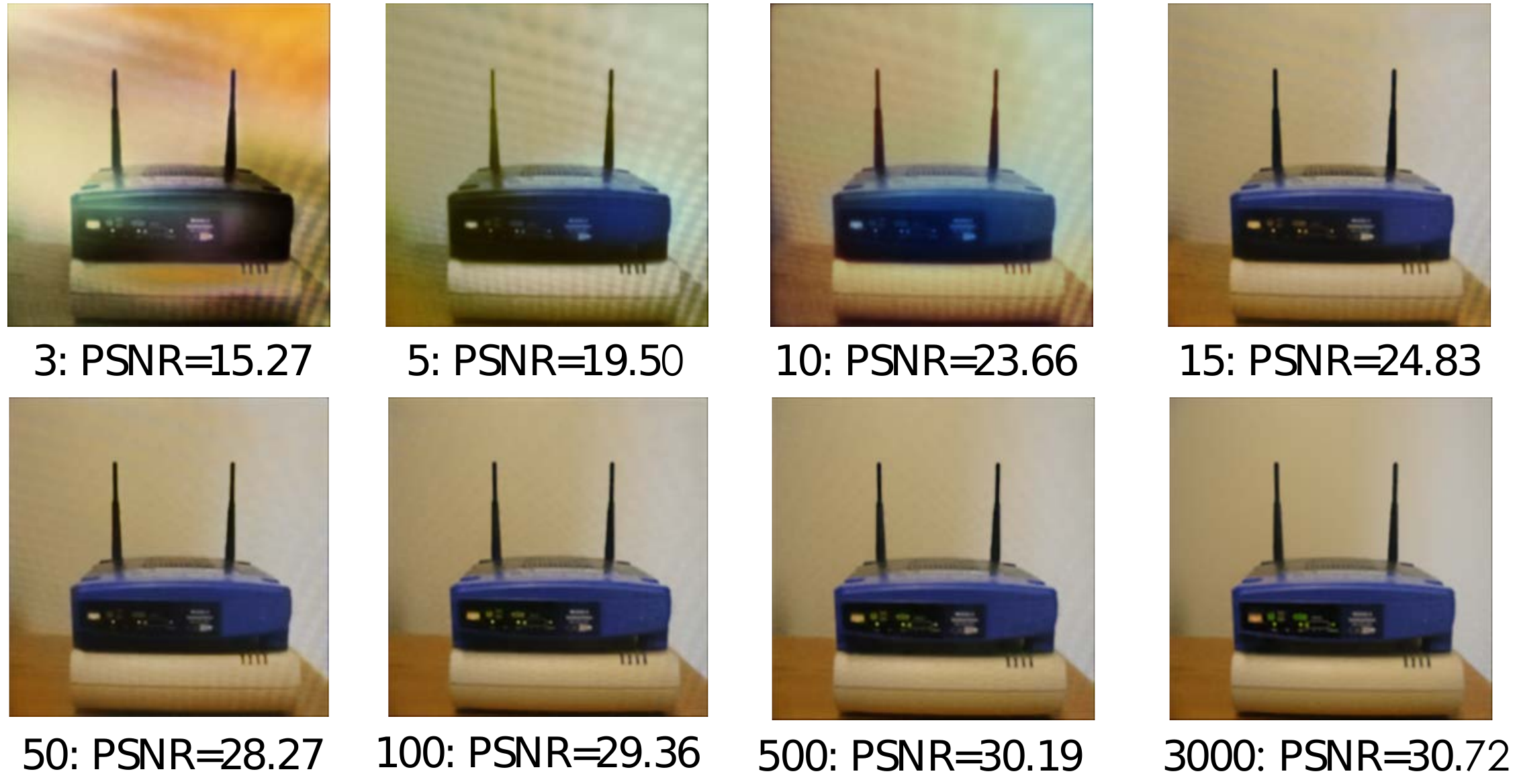}
  \caption{Intermediate results of one example in \textit{SynScreenMoire}. The numbers to the left of the PSNR are the numbers of iterations.}
  \label{fig:iter}
\end{figure}

\begin{figure}[H]
  \centering
  \includegraphics[width=1.0\textwidth]{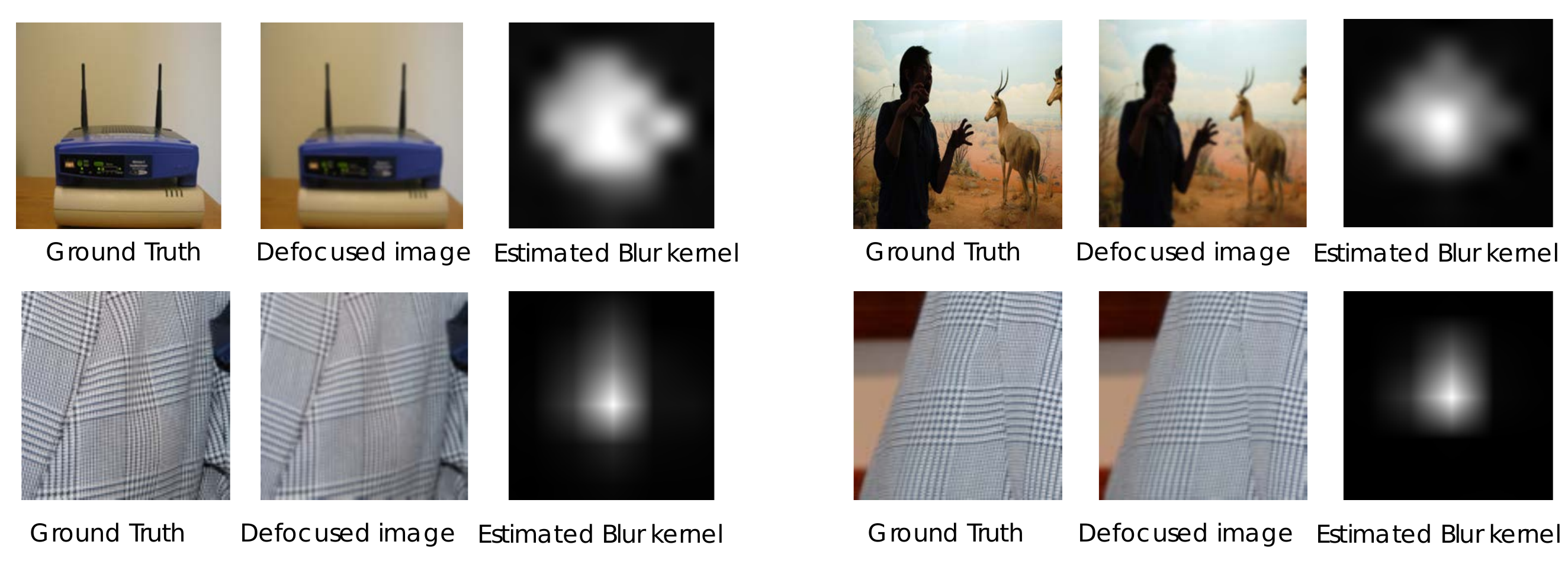}
  \caption{Visualization of estimated blur kernels.}
  \label{fig:kernel}
\end{figure}

\vspace{-0.1cm}
We visualize some intermediate results (see Figure \ref{fig:iter}), which show that as the number of iterations increases, the moir\'{e} patterns gradually disappear.
Figure \ref{fig:kernel} shows the the learnt blur kernels for different image pairs. Note that the learnt blur kernels are learned from scratch and adaptive to each image pair.

\newpage
\section{Examples of our New Dataset}
\begin{figure}[H]
  \centering
  \includegraphics[width=0.85\textwidth]{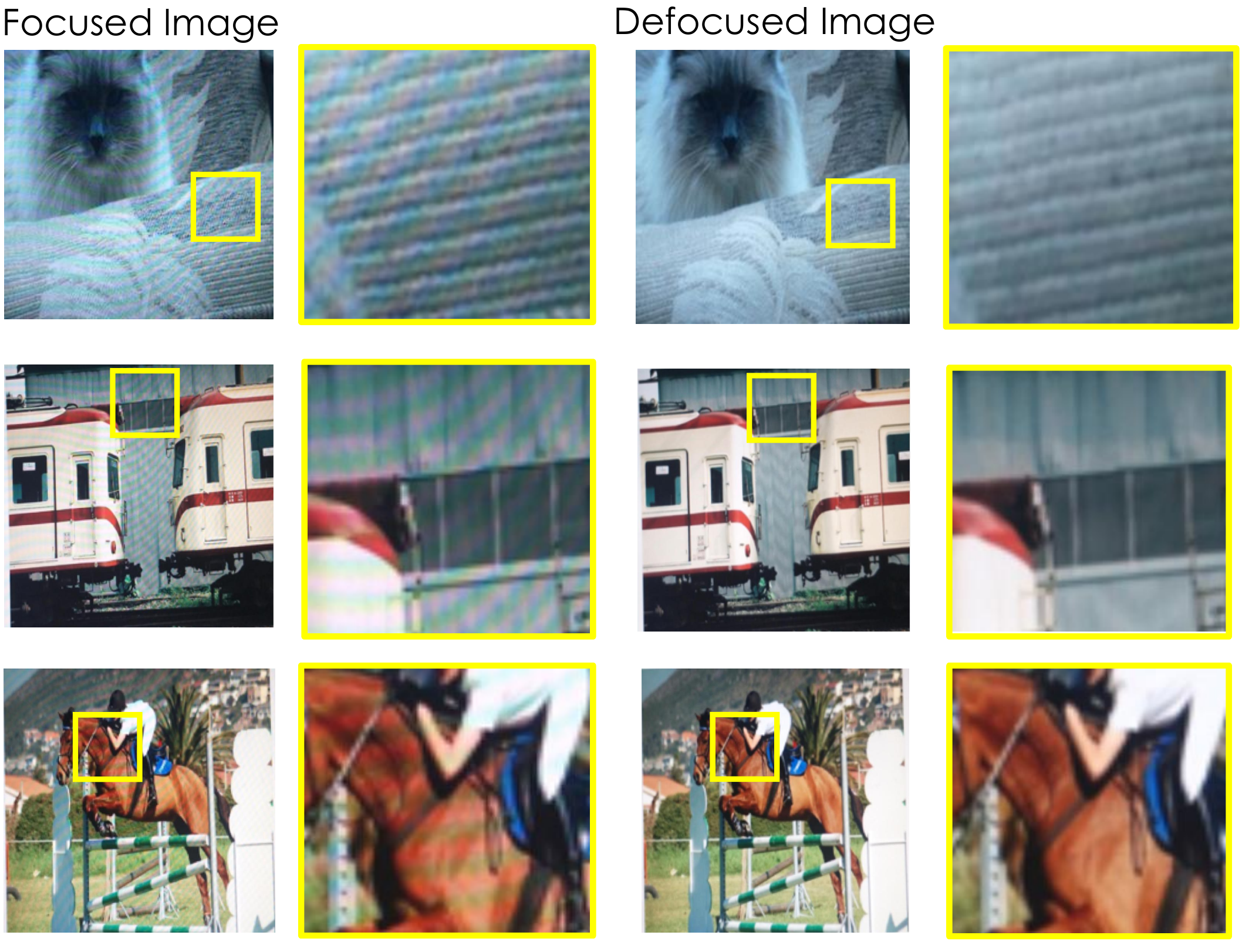}
  \caption{Examples of \textit{RealScreenMoire}.}
  \label{fig:imgsample1}
\end{figure}

\begin{figure}[H]
  \centering
  \includegraphics[width=0.85\textwidth]{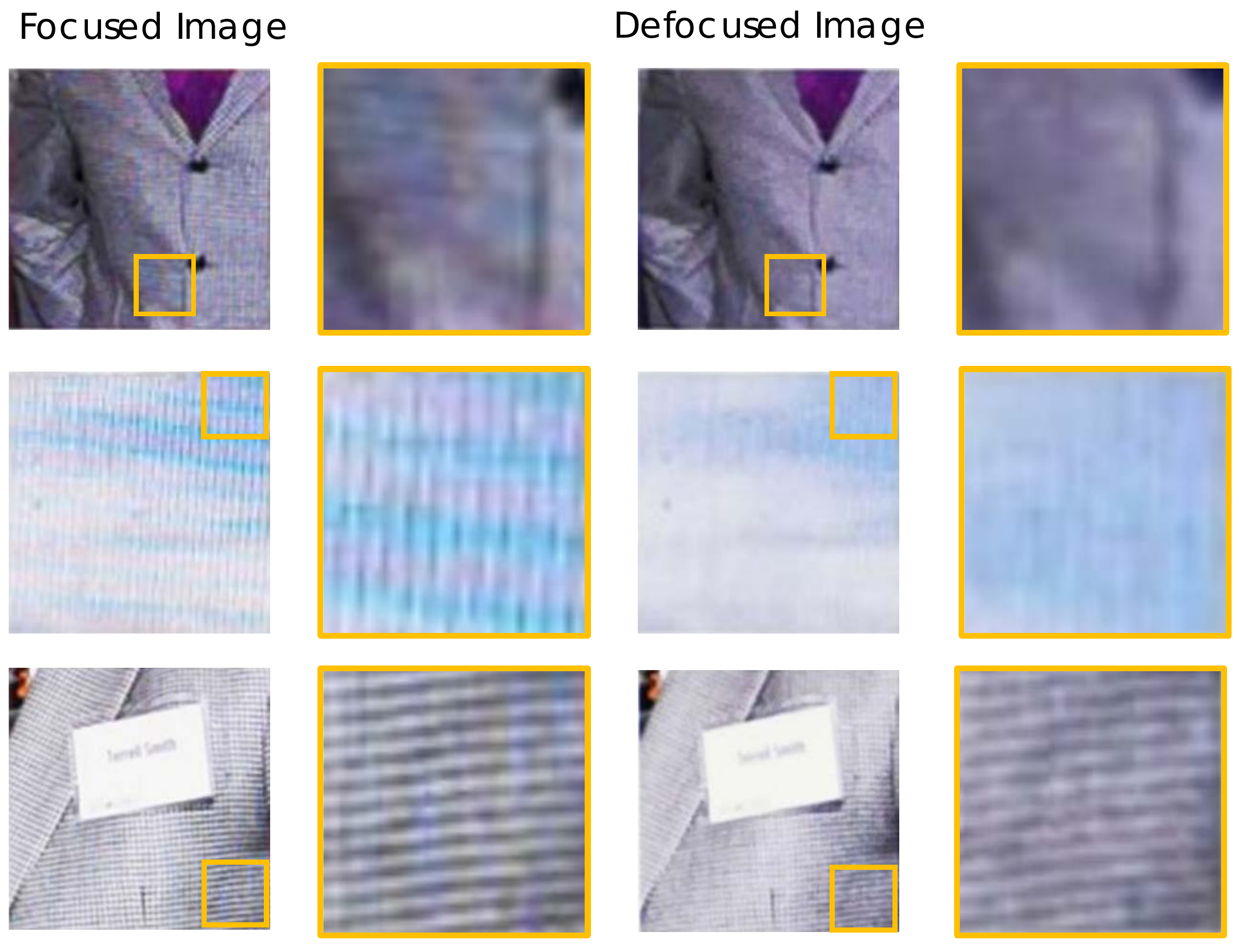}
  \caption{Examples of \textit{RealTextureMoire}.}
  \label{fig:imgsample2}
\end{figure}

%\vspace{-0.5cm}
Figures \ref{fig:imgsample1} and \ref{fig:imgsample2} show some examples of the \textit{RealScreenMoire} subset and the \textit{RealTextureMoire} subset, respectively. Note that the focused images have more details overlaid with moir\'{e} patterns, while the corresponding defocused images have no moir\'{e} patterns but appear blurry.

\section{Additional Visual Comparisons on \textit{SynScreenMoire}, \textit{SynTextureMoire}, \textit{RealScreenMoire} and \textit{RealTextureMoire}}

\begin{figure}[th]
  \centering
  \includegraphics[width=1.0\textwidth]{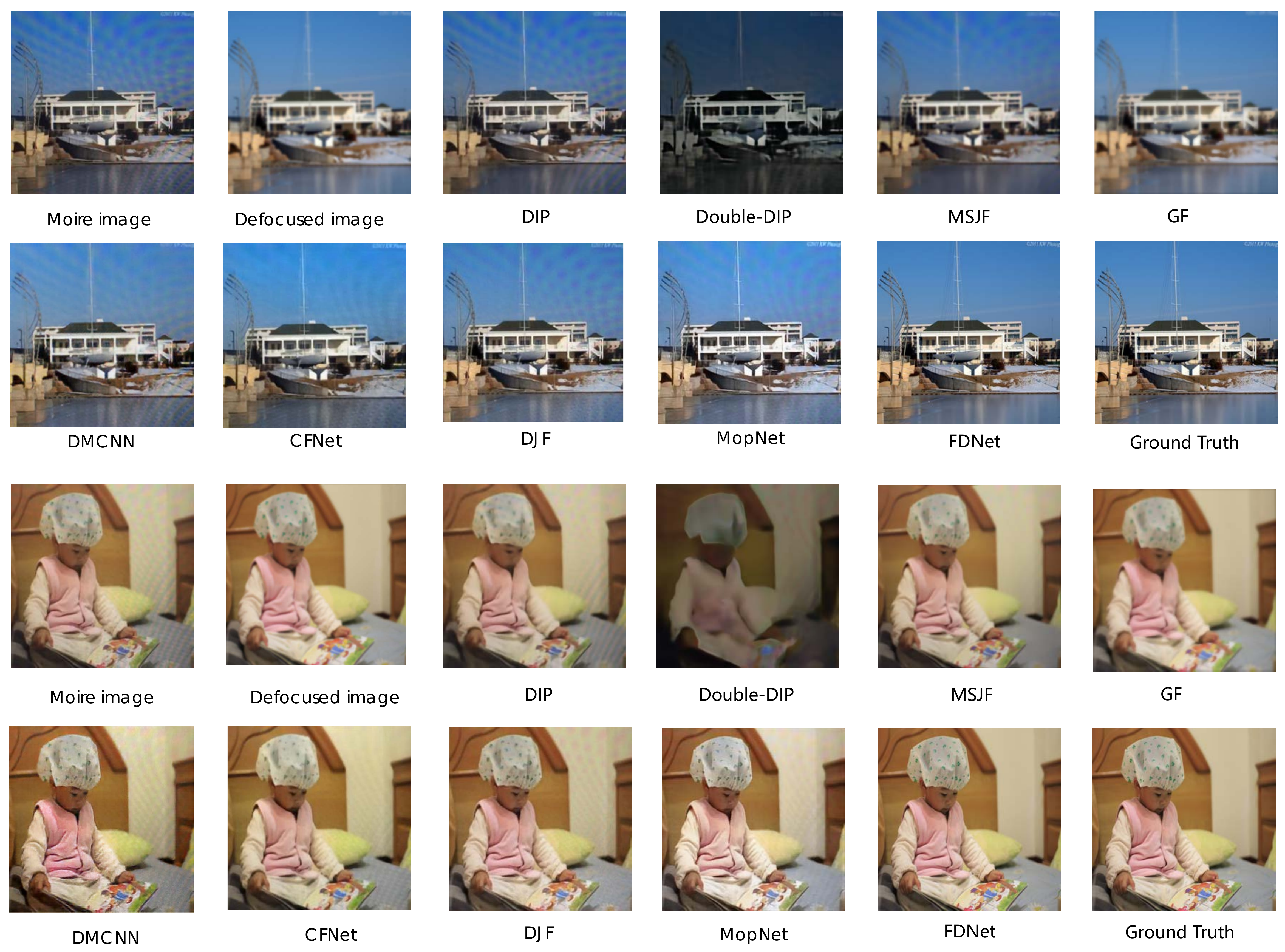}
  \caption{Visual comparisons on  \textit{SynScreenMoire}.}
  \label{fig:imgsample3}
\end{figure}

\begin{figure}[t]
  \centering
  %trim=left bottom right top, clip
  \includegraphics[trim= 2cm 0 2cm 0 clip, width=1.0\textwidth]{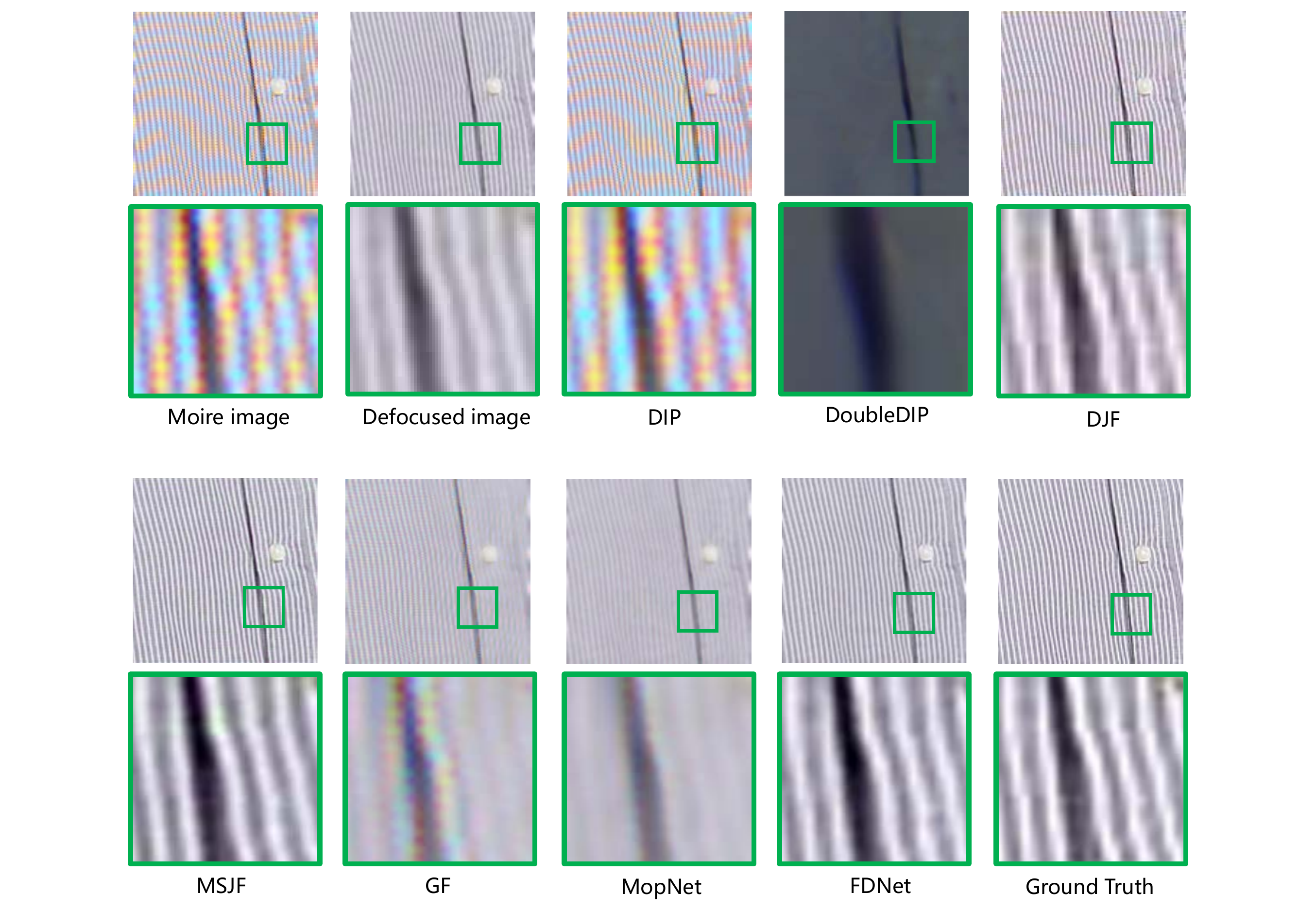}
  \caption{Visual comparisons on   \textit{SynTextureMoire}.}
  \label{fig:imgsample4}
\end{figure}

\begin{figure}[H]
  \centering
  \includegraphics[width=1.0\textwidth]{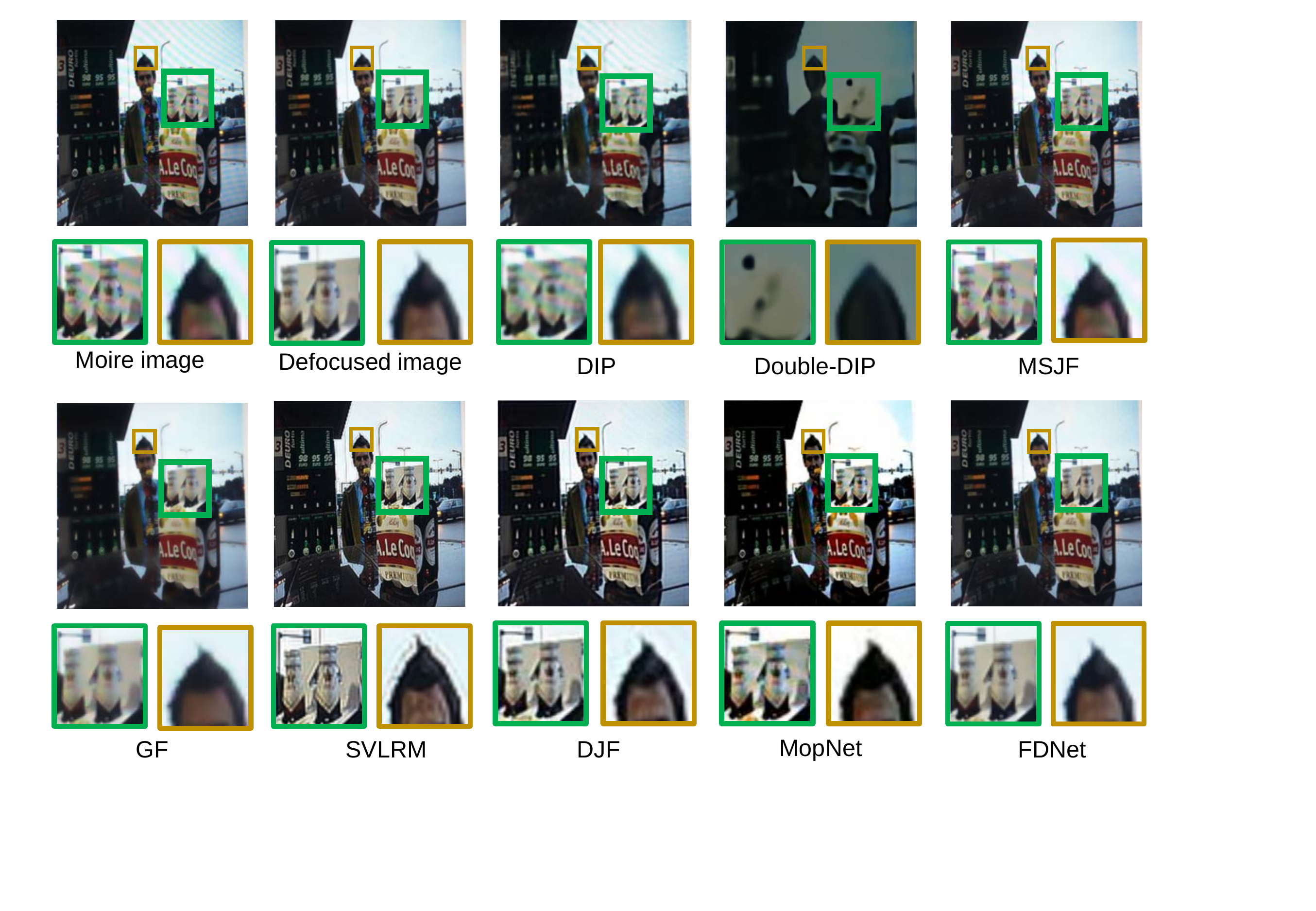}
  \caption{ Visual comparisons on  \textit{RealScreenMoire} (without ground truth).}
  \label{fig:imgsample5}
\end{figure}

\begin{figure}[H]
  \centering
  \includegraphics[width=0.9\textwidth]{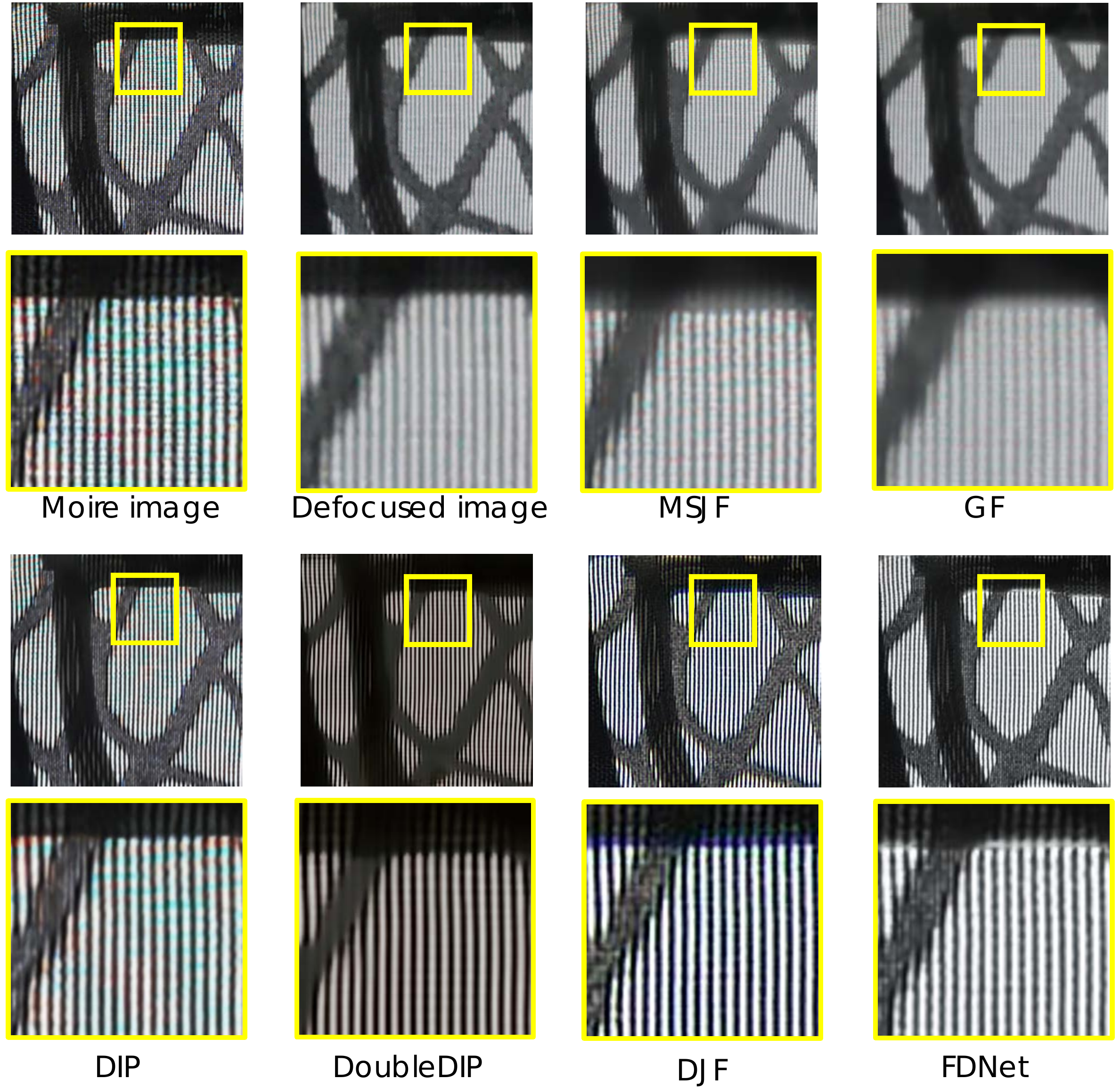}
  \caption{ Visual comparisons on  \textit{RealTextureMoire} (without ground truth).}
  \label{fig:imgsample6}
\end{figure}

Figure \ref{fig:imgsample3}, Figure \ref{fig:imgsample4}, Figure \ref{fig:imgsample5} and Figure \ref{fig:imgsample6} show more results on \textit{SynScreenMoire}, \textit{SynTextureMoire}, \textit{RealScreenMoire}, and \textit{RealScreenMoire}, respectively. The main paper presents an analysis of the results for Figures \ref{fig:imgsample3} and \ref{fig:imgsample4} on \textit{SynScreenMoire} and \textit{SynTextureMoire}.

As shown in Figures \ref{fig:imgsample5} and  \ref{fig:imgsample6}, the deep-learning based methods (SVLRM, MopNet and DJF) produce some artifacts near the edges. DJF also tends to over-sharpen the images and exhibits ringing artifacts. 
The results of DIP have obvious moir\'{e} artifacts left, and DoubleDIP has a global color shift from the original input.
In addition, the joint filtering methods (GF and MSJF) tend to smooth the high-frequency regions. 
Our FDNet outperforms all of them.

\newpage

%\section{Visual Comparisons on \textit{RealScreenMoire} and \textit{RealTextureMoire} }

\end{document}